\documentclass[twocolumn]{aastex61}
\UseRawInputEncoding
\usepackage{color}
\usepackage{natbib}

\def\star{\hbox{LTT~1445A}}

\def\mdot{\hbox{M$_{\odot}$}} 
\def\rearth{\hbox{R$_{\oplus}$}} 
\def\mearth{\hbox{M$_{\oplus}$}} 
\def\tess{\hbox{$TESS$}}

\def\pers{\hbox{s$^{-1}$}}
\def\pbmass{$2.87\pm0.25$}
\def\pbradius{$1.304^{+0.067}_{-0.060}$}
\def\pcmass{$1.54^{+0.20}_{-0.19}$}
\def\pcradius{$1.147^{+0.055}_{-0.053}$}

\providecommand{\fave}{\langle F \rangle}
\defcitealias{Winters(2019b)}{W19}

\providecommand{\bjdtdb}{\ensuremath{\rm {BJD_{TDB}}}}

\providecommand{\me}{\ensuremath{\,M_{\rm E}}}
\providecommand{\re}{\ensuremath{\,R_{\rm E}}}
\providecommand{\fave}{\langle F \rangle}
\providecommand{\fluxcgs}{10$^9$ erg s$^{-1}$ cm$^{-2}$}
\usepackage{apjfonts}

\shorttitle{A Second Planet Transiting LTT~1445A}
\shortauthors{Winters et al.}

\begin{document}

\title{A Second Planet Transiting LTT~1445A and a Determination of the Masses of Both Worlds}

\correspondingauthor{Jennifer G.\ Winters}
\email{jennifer.winters@cfa.harvard.edu}

\author[0000-0001-6031-9513]{Jennifer G.\ Winters}
\affil{Center for Astrophysics $\vert$ Harvard \& Smithsonian, 60 Garden Street, Cambridge, MA 02138, USA}
  
\author[0000-0001-5383-9393]{Ryan Cloutier}
\altaffiliation{Banting Fellow}
\affil{Center for Astrophysics $\vert$ Harvard \& Smithsonian, 60 Garden Street, Cambridge, MA 02138, USA}

\author[0000-0001-8726-3134]{Amber A.\ Medina}
\affil{Center for Astrophysics $\vert$ Harvard \& Smithsonian, 60 Garden Street, Cambridge, MA 02138, USA}

\author{Jonathan M.\ Irwin}
\affil{Center for Astrophysics $\vert$ Harvard \& Smithsonian, 60 Garden Street, Cambridge, MA 02138, USA}

\author[0000-0002-9003-484X]{David Charbonneau}
\affil{Center for Astrophysics $\vert$ Harvard \& Smithsonian, 60 Garden Street, Cambridge, MA 02138, USA}

\author[0000-0002-8462-515X]{Nicola Astudillo-Defru}
\affil{Departamento de Matem\'atica y F\'isica Aplicadas, Universidad Cat\'olica de la Sant\'isima Concepci\'on, Alonso de Rivera 2850, Concepci\'on, Chile}

\author{Xavier Bonfils}
\affil{Universit\'e Grenoble Alpes, CNRS, IPAG, F-38000 Grenoble, France}

\author[0000-0001-8638-0320]{Andrew W.\ Howard}
\affil{Department of Astronomy, California Institute of Technology, Pasadena, CA 91125, USA}

\author[0000-0002-0531-1073]{Howard Isaacson}
\affiliation{Department of Astronomy,  University of California Berkeley, Berkeley CA 94720, USA}
\affiliation{Centre for Astrophysics, University of Southern Queensland, Toowoomba, QLD, Australia}

\author[0000-0003-4733-6532]{Jacob L.\ Bean}
\affil{Department of Astronomy \& Astrophysics, University of Chicago, 5640 S. Ellis Avenue, Chicago, IL 60637, USA}

\author[0000-0003-4526-3747]{Andreas Seifahrt}
\affil{Department of Astronomy \& Astrophysics, University of Chicago, 5640 S. Ellis Avenue, Chicago, IL 60637, USA}

\author{Johanna K.\ Teske}
\affil{Earth \& Planets Laboratory of the Carnegie Institution for Science, 5241 Broad Branch Road, NW, Washington, DC 20015, USA}

\author[0000-0003-3773-5142]{Jason D.\ Eastman}
\affil{Center for Astrophysics $\vert$ Harvard \& Smithsonian, 60 Garden Street, Cambridge, MA 02138, USA}

\author[0000-0002-6778-7552]{Joseph D.\ Twicken}
\affil{SETI Institute, Moffett Field, CA 94035, USA}
\affil{NASA Ames Research Center, Moffett Field, CA 94035, USA}

\author[0000-0001-6588-9574]{Karen A.\ Collins}
\affil{Center for Astrophysics $\vert$ Harvard \& Smithsonian, 60 Garden Street, Cambridge, MA 02138, USA}

\author[0000-0002-4625-7333]{Eric L.\ N.\ Jensen}
\affiliation{Department of Physics \& Astronomy, Swarthmore College, Swarthmore PA 19081, USA}

\author[0000-0002-8964-8377]{Samuel N.\ Quinn}
\affil{Center for Astrophysics $\vert$ Harvard \& Smithsonian, 60 Garden Street, Cambridge, MA 02138, USA}

\author[0000-0001-5133-6303]{Matthew J.\ Payne}
\affil{Center for Astrophysics $\vert$ Harvard \& Smithsonian, 60 Garden Street, Cambridge, MA 02138, USA}

\author[0000-0002-2607-138X]{Martti H.\ Kristiansen}
\affiliation{Brorfelde Observatory, Observator Gyldenkernes Vej 7, DK-4340 T\o{}ll\o{}se, Denmark}
\affiliation{DTU Space, National Space Institute, Technical University of Denmark, Elektrovej 327, DK-2800 Lyngby, Denmark}

\author{Alton Spencer}
\affil{Western Connecticut State University, Danbury CT 06810, USA}

\author[0000-0001-7246-5438]{Andrew Vanderburg}
\affiliation{Department of Physics and Kavli Institute for Astrophysics and Space Research, Massachusetts Institute of Technology, Cambridge, MA 02139, USA} 

\author[0000-0002-6532-4378]{Mathias Zechmeister}
\affil{Institut f\"ur Astrophysik, Georg-August-Universit\"at,
Friedrich- Hund-Platz 1, D-37077 G\"ottingen, Germany}

\author[0000-0002-3725-3058]{Lauren M. Weiss}
\affiliation{Department of Physics, University of Notre Dame, Nieuwland Science Hall, Notre Dame, IN 46556, USA}

\author[0000-0002-6937-9034]{Sharon Xuesong Wang}
\affiliation{Department of Astronomy, Tsinghua University, Beijing 100084, People's Republic of China}

\author[0000-0003-3092-4418]{Gavin Wang}
\affiliation{Tsinghua International School, Beijing 100084, China}
\affiliation{Stanford Online High School, 415 Broadway Academy Hall, Floor 2, 8853, Redwood City, CA 94063, USA}

\author{St\'ephane Udry}
\affil{Observatoire de Gen\`eve, Universit\'e de Gen\`eve, Chemin Pegasi 51, 1290 Sauverny, Switzerland}

\author[0000-0002-0654-4442]{Ivan A. Terentev}  
\affiliation{Citizen Scientist, Planet Hunter, Petrozavodsk, Russia}

\author[0000-0002-4410-4712]{Julian St\"urmer} 
\affil{Landessternwarte, Zentrum f\"ur Astronomie der Universit\"at Heidelberg, K\"onigstuhl 12, D-69117 Heidelberg, Germany}

\author[0000-0001-7409-5688]{Guðmundur Stef\'ansson}
\affil{Princeton University, Department of Astrophysical
Sciences, 4 Ivy Lane, Princeton, NJ 08540, USA}
\altaffiliation{Henry Norris Russell Fellow}

\author[0000-0002-1836-3120]{Avi~Shporer}
\affiliation{Department of Physics and Kavli Institute for Astrophysics and Space Research, Massachusetts Institute of Technology, Cambridge, MA 02139, USA}

\author{Stephen Shectman}
\affil{The Observatories of the Carnegie Institution for Science, 813 Santa Barbara Street, Pasadena, CA 91101, USA}

\author[0000-0003-3904-6754]{Ramotholo Sefako}
\affiliation{South African Astronomical Observatory, P.O. Box 9, Observatory, Cape Town 7935, South Africa}

\author[0000-0002-1637-2189]{Hans Martin Schwengeler}
\affiliation{Citizen Scientist, Planet Hunter, Bottmingen, Switzerland}

\author[0000-0001-8227-1020]{Richard P. Schwarz}
\affiliation{Patashnick Voorheesville Observatory, Voorheesville, NY 12186, USA}

\author[0000-0003-3623-7280]{Nicholas Scarsdale}
\affiliation{Department of Astronomy and Astrophysics, University of California, Santa Cruz, CA 95060, USA}

\author[0000-0003-3856-3143]{Ryan A. Rubenzahl}
\altaffiliation{NSF Graduate Research Fellow}
\affiliation{Department of Astronomy, California Institute of Technology, Pasadena, CA 91125, USA}

\author[0000-0001-8127-5775]{Arpita Roy}
\affiliation{Space Telescope Science Institute, 3700 San Martin Drive, Baltimore, MD 21218, USA}
\affiliation{Department of Physics and Astronomy, Johns Hopkins University, 3400 N Charles St, Baltimore, MD 21218, USA}

\author{Lee J.\ Rosenthal}
\affiliation{Department of Astronomy, California Institute of Technology, Pasadena, CA 91125, USA}

\author[0000-0003-0149-9678]{Paul Robertson}
\affiliation{Department of Physics \& Astronomy, University of California Irvine, Irvine, CA 92697, USA}

\author[0000-0003-0967-2893]{Erik A. Petigura}
\affiliation{Department of Physics \& Astronomy, University of California Los Angeles, Los Angeles, CA 90095, USA}

\author{Francesco Pepe}
\affil{Observatoire de Gen\`eve, Universit\'e de Gen\`eve, Chemin Pegasi 51, 1290 Sauverny, Switzerland}

\author{Mark Omohundro}
\affiliation{Citizen Scientist, c/o Zooniverse, Department of Physics, University of Oxford, Denys Wilkinson Building, Keble Road, Oxford, OX1 3RH, UK}

\author[0000-0001-8898-8284]{Joseph M. Akana Murphy}
\altaffiliation{NSF Graduate Research Fellow}
\affiliation{Department of Astronomy and Astrophysics, University of California, Santa Cruz, CA 95064, USA}

\author[0000-0001-9087-1245]{Felipe Murgas}
\affil{Instituo de Astrof\'isica da Canarias (IAC), 38205 La Laguna, Tenerife, Spain}
\affil{Departamento de Astrof\'isica, Universidad de La Laguna (ULL), E-38206 La Laguna, Tenerife, Spain} 

\author[0000-0003-4603-556X]{Teo Mo\v{c}nik}
\affiliation{Gemini Observatory/NSF's NOIRLab, 670 N. A'ohoku Place, Hilo, HI 96720, USA}

\author[0000-0001-7516-8308]{Benjamin T. Montet}
\affil{School of Physics, University of New South Wales, Sydney, NSW 2052, Australia}
\affil{UNSW Data Science Hub, University of New South Wales, Sydney, NSW 2052, Australia}

\author[0000-0002-6245-0264]{Ronald Mennickent}
\affil{Departamento de Astronom{\'{i}}a, Universidad de Concepci\'on, Casilla 160-C, Concepci\'on, Chile}

\author[0000-0002-7216-2135]{Andrew W. Mayo}
\affil{Department of Astronomy, University of California Berkeley, Berkeley, CA 94720, USA}
\affil{Centre for Star and Planet Formation, Natural History Museum of Denmark \& Niels Bohr Institute, University of Copenhagen, \O ster Voldgade 5-7, DK-1350 Copenhagen K., Denmark}

\author[0000-0001-8879-7138]{Bob Massey}
\affil{Villa '39 Observatory, Landers, CA 92285, USA}

\author[0000-0001-8342-7736]{Jack Lubin}
\affiliation{Department of Physics \& Astronomy, University of California Irvine, Irvine, CA 92697, USA}

\author{Christophe Lovis}
\affil{Observatoire de Gen\`eve, Universit\'e de Gen\`eve, Chemin Pegasi 51, 1290 Sauverny, Switzerland}

\author[0000-0003-0828-6368]{Pablo Lewin}
\affiliation{The Maury Lewin Astronomical Observatory, Glendora, CA 91741, USA}

\author[0000-0003-0534-6388]{David Kasper}
\affil{Department of Astronomy \& Astrophysics, University of Chicago, 5640 S. Ellis Avenue, Chicago, IL 60637, USA}

\author[0000-0002-7084-0529]{Stephen R. Kane} 
\affiliation{Department of Earth and Planetary Sciences, University of California, Riverside, CA 92521, USA}

\author[0000-0002-4715-9460]{Jon M.\ Jenkins}
\affil{NASA Ames Research Center, Moffett Field, CA 94035, USA}

\author[0000-0001-8832-4488]{Daniel Huber}
\affiliation{Institute for Astronomy, University of Hawai`i, 2680 Woodlawn Drive, Honolulu, HI 96822, USA}

\author[0000-0003-1728-0304]{Keith Horne}
\affiliation{SUPA Physics and Astronomy, University of St. Andrews, Fife, KY16 9SS Scotland, UK}

\author[0000-0002-0139-4756]{Michelle L.\ Hill}
\affiliation{Department of Earth and Planetary Sciences, University of California, Riverside, CA 92521, USA}

\author[0000-0002-6394-6544]{Paula Gorrini}
\affil{Institut f\"ur Astrophysik, Georg-August-Universit\"at,
Friedrich- Hund-Platz 1, D-37077 G\"ottingen, Germany}

\author[0000-0002-8965-3969]{Steven Giacalone}
\affil{Department of Astronomy, University of California Berkeley, Berkeley, CA 94720, USA}

\author[0000-0003-3504-5316]{Benjamin Fulton}
\affiliation{NASA Exoplanet Science Institute/Caltech-IPAC, MC 314-6, 1200 E. California Blvd., Pasadena, CA 91125, USA}

\author[0000-0003-0536-4607]{Thierry Forveille}
\affil{Universit\'e Grenoble Alpes, CNRS, IPAG, F-38000 Grenoble, France}

\author{Pedro Figueira}
\affil{European Southern Observatory, Alonso de C\'ordova 3107, Vitacura, Regi\'on Metropolitana, Chile}
\affil{Instituto de Astrof\'isica e Ci\^{e}ncias do Espa\c{c}o, Universidade do Porto, CAUP, Rua das Estrelas, 4150-762 Porto, Portugal}

\author[0000-0002-3551-279X]{Tara Fetherolf}
\affiliation{Earth and Planetary Sciences Department, University of California, Riverside, CA 92521, USA}
\altaffiliation{UC Chancellor's Fellow}

\author[0000-0001-8189-0233]{Courtney Dressing}
\affiliation{Department of Astronomy,  University of California Berkeley, Berkeley CA 94720, USA}

\author[0000-0001-9289-5160]{Rodrigo F.\ D\'iaz}
\affil{International Center for Advanced Studies (ICAS) and ICIFI (CONICET), ECyT-UNSAM, Campus Miguelete, 25 de Mayo y Francia, (1650) Buenos Aires, Argentina}

\author[0000-0001-5099-7978]{Xavier Delfosse}
\affil{Universit\'e Grenoble Alpes, CNRS, IPAG, F-38000 Grenoble, France}

\author[0000-0002-4297-5506]{Paul A.\ Dalba}
\altaffiliation{NSF Astronomy and Astrophysics Postdoctoral Fellow}
\affiliation{Department of Astronomy and Astrophysics, University of California, Santa Cruz, CA 95064, USA}
\affiliation{Department of Earth and Planetary Sciences, University of California, Riverside, CA 92521, USA}

\author[0000-0002-8958-0683]{Fei Dai}
\affiliation{Division of Geological and Planetary Sciences, California Institute of Technology, Pasadena, CA 91125, USA}

\author{C.C.Cort\'es} 
\affil{Departamento de F\'isica, Facultad de Ciencias, Universidad del B\'io-B\'io, Avenida Collao 1202, Casilla 15-C, Concepci\'on, Chile}
\affil{Departamento de Astronom\'ia, Universidad de Concepci\'on, Casilla 160-C, Concepci\'on, Chile}

\author{Ian J.\ M.\ Crossfield}
\affiliation{Department of Physics \& Astronomy, University of Kansas, 1082 Malott, 1251 Wescoe Hall Dr., Lawrence, KS 66045, USA}

\author[0000-0002-5226-787X]{Jeffrey~D.~Crane}
\affil{The Observatories of the Carnegie Institution for Science, 813 Santa Barbara Street, Pasadena, CA 91101, USA}

\author[0000-0003-2239-0567]{Dennis M.\ Conti}
\affiliation{American Association of Variable Star Observers, 49 Bay State Road, Cambridge, MA 02138, USA}

\author[0000-0003-2781-3207]{Kevin I.\ Collins}
\affiliation{George Mason University, 4400 University Drive, Fairfax, VA 22030 USA}

\author[0000-0003-1125-2564]{Ashley Chontos}
\altaffiliation{NSF Graduate Research Fellow}
\affiliation{Institute for Astronomy, University of Hawai`i, 2680 Woodlawn Drive, Honolulu, HI 96822, USA}

\author[0000-0003-1305-3761]{R.~Paul~Butler}
\affil{Earth \& Planets Laboratory, Carnegie Institution for Science, 5241 Broad Branch Road, NW, Washington, DC 20015, USA}

\author[0000-0002-3481-9052]{Peyton Brown}
\affiliation{Department of Physics and Astronomy, Vanderbilt University, 6301 Stevenson Center Ln., Nashville, TN 37235, USA}

\author{Madison Brady}
\affil{Department of Astronomy \& Astrophysics, University of Chicago, 5640 S. Ellis Avenue, Chicago, IL 60637, USA}


\author[0000-0003-0012-9093]{Aida Behmard}
\altaffiliation{NSF Graduate Research Fellow}
\affiliation{Division of Geological and Planetary Sciences, California Institute of Technology, Pasadena, CA 91125, USA}

\author[0000-0001-7708-2364]{Corey Beard}
\affiliation{Department of Physics \& Astronomy, University of California Irvine, Irvine, CA 92697, USA}

\author[0000-0002-7030-9519]{Natalie M. Batalha}
\affiliation{Department of Astronomy and Astrophysics, University of California, Santa Cruz, CA 95060, USA}

\author[0000-0003-3208-9815]{Jose-Manuel Almenara}
\affil{Universit\'e Grenoble Alpes, CNRS, IPAG, F-38000 Grenoble, France}

\begin{abstract}

LTT~1445 is a hierarchical triple M-dwarf star system located at a distance of 6.86 parsecs. The primary star LTT~1445A (0.257 M$_\Sun$) is known to host the transiting planet LTT~1445Ab with an orbital period of 5.36~days, making it the second closest known transiting exoplanet system, and the closest one for which the host is an M dwarf.  Using {\it TESS} data, we present the discovery of a second planet in the LTT~1445 system, with an orbital period of 3.12~days. We combine radial velocity measurements obtained from the five spectrographs ESPRESSO, HARPS, HIRES, MAROON-X, and PFS to establish that the new world also orbits LTT~1445A. We determine the mass and radius of LTT~1445Ab to be \pbmass ~\mearth ~and \pbradius ~\rearth, consistent with an Earth-like composition. For the newly discovered LTT~1445Ac, we measure a mass of \pcmass ~\mearth ~and a minimum radius of 1.15 \rearth, but we cannot determine the radius directly as the signal-to-noise of our light curve permits both grazing and non-grazing configurations. Using MEarth photometry and ground-based spectroscopy, we establish that star C (0.161 M$_\Sun$) is likely the source of the 1.4-day rotation period, and star B (0.215 M$_\Sun$) has a likely rotation period of 6.7~days. Although we have not yet determined the rotation period of star A, we estimate a probable rotation period of 85~days. Thus, this triple M-dwarf system appears to be in a special evolutionary stage where the most massive M dwarf has spun down, the intermediate mass M dwarf is in the process of spinning down, while the least massive stellar component has not yet begun to spin down.  

\end{abstract}

\section{Introduction}

Terrestrial exoplanets are difficult to study. Whether one is seeking to measure their bulk densities, search for attendant satellites, or probe their atmospheres, the task is made easier by two factors: First, the stars must be nearby, so that we might mitigate the cruel dictates of the inverse square law. Second, the stellar mass and radius should be small, so that the ratio of the planetary and stellar quantities is maximized. It is for these reasons that nearby M-dwarf systems are the darlings of the exoplanet menagerie. Fortunately, M dwarfs are abundant, making up approximately 72\% of all stars in the stellar neighborhood \citep{Henry(2018)}. Their dominance among nearby planetary hosts is likewise in evidence: Indeed, of the 16 stars within 15 parsecs that are known to host transiting exoplanets, all but 4 are M dwarfs.

At a distance of 6.53~pc, the closest star known to host transiting exoplanets is HD~219134 \citep{Motalebi(2015), Gillon(2017)}. Although the two transiting worlds have had their densities measured, they are otherwise largely inaccessible because the parent star is a K dwarf with a radius of 0.778~R$_\Sun$. But the star in second place is much more diminutive in stature: LTT~1445A, with a radius of 0.268~R$_\Sun$ and a mass of 0.257~M$_\Sun$. At a distance of 6.86~pc \citep{Gaia(2016a),Lindegren(2021)}, it is the closest M dwarf known to host a transiting planet.  Using data from the {\it Transiting Exoplanet Survey Satellite} ($TESS$) \citep{Ricker(2015)} and a host of ground-based telescopes, \citet[][hereafter \citetalias{Winters(2019b)}]{Winters(2019b)}, published the discovery of LTT~1445Ab, a planet with an orbital period of 5.36~days. Although that planet was presumed to be terrestrial owing to its radius, no measurement of its mass has yet been presented. A mass measurement is crucial for at least two reasons: First, it would constrain the bulk composition of the planet. If the planet proves to be terrestrial, the mass measurement would allow for the determination of the ratio of its iron-and-nickel core to its magnesium-silicate mantle, which in turn could be compared to elemental abundances of the stellar photosphere. Second, future atmospheric studies require a direct estimate of the surface gravity to enable an unambiguous interpretation of data gathered in transmission, since the atmospheric scale height is a combination of the surface gravity, temperature, and mean molecular weight. Intriguingly, LTT~1445A is the most massive member of a resolved triple M-dwarf system. The three stars fall on a line (see Fig.~1 from \citetalias{Winters(2019b)}) and the stellar orbital plane of stars B and C is viewed edge-on, and hence the entire system may be co-planar.

In pursuit of the mass of LTT~1445Ab, we set out to gather multi-epoch radial velocity (RV) measurements with five high-resolution spectrographs. While that activity was underway, we found a second periodic transit-like signal in the {\it TESS} data. Combining the RV and {\it TESS} data with ground-based photometric monitoring, we established that indeed a second planet transits LTT~1445A, and orbits interior to the known world. In what follows, we describe these findings, along with our measurements of the masses of both worlds. 

We briefly summarize the host star system in \S \ref{sec:host_system}, before turning to a description of our photometric and spectroscopic data in \S \ref{sec:obs}. We present our analysis of these data in \S \ref{sec:analysis}, and their implications in \S \ref{sec:discuss}. 

\section{Description of the Host Stellar System}\label{sec:host_system}

The host system, LTT~1445ABC\footnote{Other names for A: TIC~98796344, TOI~455, L~730-18, BD-17~588A, RST~2292A, WDS~J03019-1633A,  2MASS~J03015142-1635356, Gaia~DR2~5153091836072107136, GJ~3193; other names for BC: TIC~98796342, BD-17~588B, RST~2292BC, WDS~J03019-1633B, 2MASS~J03015107-1635306, Gaia~DR2~5153091836072107008.} \citep{Luyten(1957),Luyten(1980a)}, is a nearby, hierarchical trio of mid-to-late M dwarfs.  The preliminary astrometric orbit for LTT~1445BC, presented in \citetalias{Winters(2019b)}, indicates an orbital period of $36.2 \pm 5.3$ yrs and a semi-major axis $1.159 \pm 0.076$ arcseconds, which corresponds to an average physical separation of $8.1 \pm 0.5$ AU. As noted in \citetalias{Winters(2019b)}, the separation between the A and BC appears to have recently begun decreasing, with the most recent value of 7\farcs10 obtained in 2017, according to data available in the Washington Double Star (WDS) Catalog\footnote{\url{https://www.usno.navy.mil/USNO/astrometry/optical-IR-prod/wds/WDS}} \citep{Mason(2009b)}.  Considering the total mass of the three stellar components and an average angular separation of 5\arcsec ~(corresponding to 34 AU, calculated using the weighted mean of the parallaxes of A and BC: $145.6917 \pm 0.0244$ mas), we estimate the period of the A-BC orbit to be roughly 250 years. We refer the reader to Figure~1 in \citetalias{Winters(2019b)}, which presents a resolved image of the 3 stars in the system. We also refer the reader to Table 1 in  \citetalias{Winters(2019b)}, which lists the astrometry, photometry, and stellar parameters of LTT~1445.

\section{Observations}
\label{sec:obs}

\subsection{Photometric Time Series Data}

We describe here the photometric time series observations of LTT~1445 by \tess, the MEarth-South telescope array, and the Las Cumbres Observatory telescope network.

\subsubsection{TESS Observations}
\label{subsubsec:tess}

$TESS$ observed LTT~1445 in the first year of its prime mission, which resulted in the discovery of LTT~1445Ab, as described in \citetalias{Winters(2019b)}. We included this system in our \tess ~Guest Investigator program (PI Winters; G03250) target list to gather short-cadence (two-minute) data of the volume-complete sample of M dwarfs with masses $0.1 \leq {\rm M/M_{\odot}} \leq 0.3$ within 15 parsecs \citep{Winters(2021)}. LTT~1445A and BC were also included in the \tess ~Input Catalog (TIC) and Candidate Target List (CTL) \citep{Stassun(2018)} via the Cool Dwarf Sample \citep{Muirhead(2018)}. The LTT~1445 system was re-observed at two-minute cadence in sector 31 from UT 2020 October 22 to November 16, in spacecraft orbits 69 and 70. The system fell on CCD 4 of Camera 2. 

As described in the data release notes\footnote{\url{ https://archive.stsci.edu/missions/tess/doc/tess_drn/tess_sector_31_drn47_v02.pdf}} for sector 31, data collection was paused for 1.4 days between the two orbits to download data, a single momentum dump was conducted halfway through each orbit, and a star tracker anomaly caused the data collection in orbit 70 to end 2.08 days early.

The two-minute cadence data were reduced with the NASA Ames Science Processing Operations Center (SPOC) pipeline \citep{Jenkins(2015),Jenkins(2016)} that was repurposed from the $Kepler$ reduction pipeline \citep{Jenkins(2010)}. A planet candidate with an approximate radius of 1.5$\pm$0.4 \rearth ~was again detected in four transits during sector 31 to have a period of 5.36 days and a transit depth of 2750$\pm$191 ppm with a signal-to-noise ratio of 15.2. 

A second planet with a 3.12-day period and a transit duration of roughly 45 min was identified in the sector 4 light curve of LTT~1445 by several of us (MHK, IAT, HMS, MO, AS). After the release of the sector 31 light curve data, we recovered a very shallow transit signal with a period of P = 3.12 days. This second planet has not been identified by the $TESS$ planet-vetting team, and thus is not assigned a TOI designation; however, the $TESS$ SPOC algorithm recovered signals at periods of 1.5623 days and 1.5616 days, half that of the expected period, with a multiple event statistic \citep[MES,][equivalent to the signal to noise of the transit in the folded light curve]{Jenkins(2002)} of 6.77 and 5.36 for the sector 4 and 31 data, respectively. To advance to the vetting process, planet candidates must have a SPOC MES larger than 7.1. The SPOC team manually re-ran their data validation algorithm \citep{Twicken(2018),Li(2019)}, allowing for both the 1.56- and 3.12-day periods. While the 1.56-day signal was strong, the candidate failed the odd/even transit test at this period. The 3.12-day signal passed the odd/even transit test. The 3.12-day signal consisted of six observed transits in the sector 4 light curve yielding a transit depth of $1554\pm187$ ppm, and eight transits in the sector 31 light curve, yielding a transit depth of $1607\pm192$ ppm. However, the SNR of the light curve was insufficient to use the $TESS$ pixel-level diagnostics to identify which star in LTT~1445 was the origin of the 3.12-day transit signal. 


There is no background star from which the transit could originate. \citetalias{Winters(2019b)} compared the position of the system in archival images taken in 1953 to its position at the time the TESS data were taken in 2018. The system's high proper motion (roughly half an arcsecond per year) and the sixty-five year difference in the images permitted \citetalias{Winters(2019b)} to eliminate the hypothesis that the transit signal originates from a background star coincident with the position of LTT1445 at the $TESS$ epochs. The system has moved only roughly 1\arcsec~between the 2018 and 2020 data, so this conclusion remains valid. We are thus confident that the transit signal originates in the LTT~1445 system.

\subsubsection{MEarth-South Photometric Monitoring}
\label{subsubsec:mearth_south_photometric_monitoring}

For the purpose of determining the rotation periods of A and BC, we gathered photometric monitoring observations using MEarth-South
\citep{Nutzman(2008),Irwin(2015)} telescope 6 at Cerro Tololo Inter-American Observatory (CTIO), Chile,
beginning UT 2019 February 14; observations are ongoing at the time of writing.
The dataset used in the present analysis includes 330 nights of data
ending 2021 October 7.  We gathered no observations between 2020
March 10 and 2020 November 4, when CTIO was closed due to the COVID-19 pandemic.

We gathered data in visits of $5 \times 4$ seconds at a cadence of
approximately 20 minutes between visits while the target was above 2
airmasses and there were no higher priority observations such as
transit followup.  We obtained a total of 17,987 exposures.

The delivered image quality of the MEarth telescopes can be highly
variable over long time baselines, predominantly due to wind shake. Wind at the location of MEarth-South on Cerro Tololo shows a
seasonal variation, with stronger and more gusty winds during the
winter and spring seasons in particular.  FWHM values ranging from
approximately 1.7 to 5.3 pixels were recorded during the time series,
where the pixel scale is $0.84$ arcsec/pixel.  Consequently while the
BC pair is always blended, there is variable blending between A and
BC, requiring specialized photometric extraction procedures.

As described in \citetalias{Winters(2019b)}, separate apertures of 4.2
pixel radius were placed on the A and BC components, using a global
(full-field) astrometric solution for aperture positioning to avoid
drift in the relative positions of the A and BC apertures as a
function of image quality.  For the present purpose we require
separate light curves of A and BC, so external comparison stars
elsewhere in the field were used rather than the procedure used for
the transit followup observations in our previous work.  The LTT~1445
system is in a sparse field so there were relatively few of these
available, and they are fainter than the target stars. Thus, the
resulting photometric solutions exhibit higher noise levels and systematic errors than are typical for
MEarth data.

The comparison star solutions are unable to correct the effect of
variable blending between A and BC, which causes fairly severe
systematic errors in the resulting light curves.  In order to partly
correct this effect we modified our standard techniques used for
periodogram analysis of MEarth data detailed in
\citet{Newton(2016)} to use simultaneous linear decorrelation
against both the standard ``common mode'' term (to correct variable
atmospheric precipitable water vapor) and against full-width-half-maximum (FWHM).

\begin{figure*}
\centering
\includegraphics[scale=.60,angle=0]{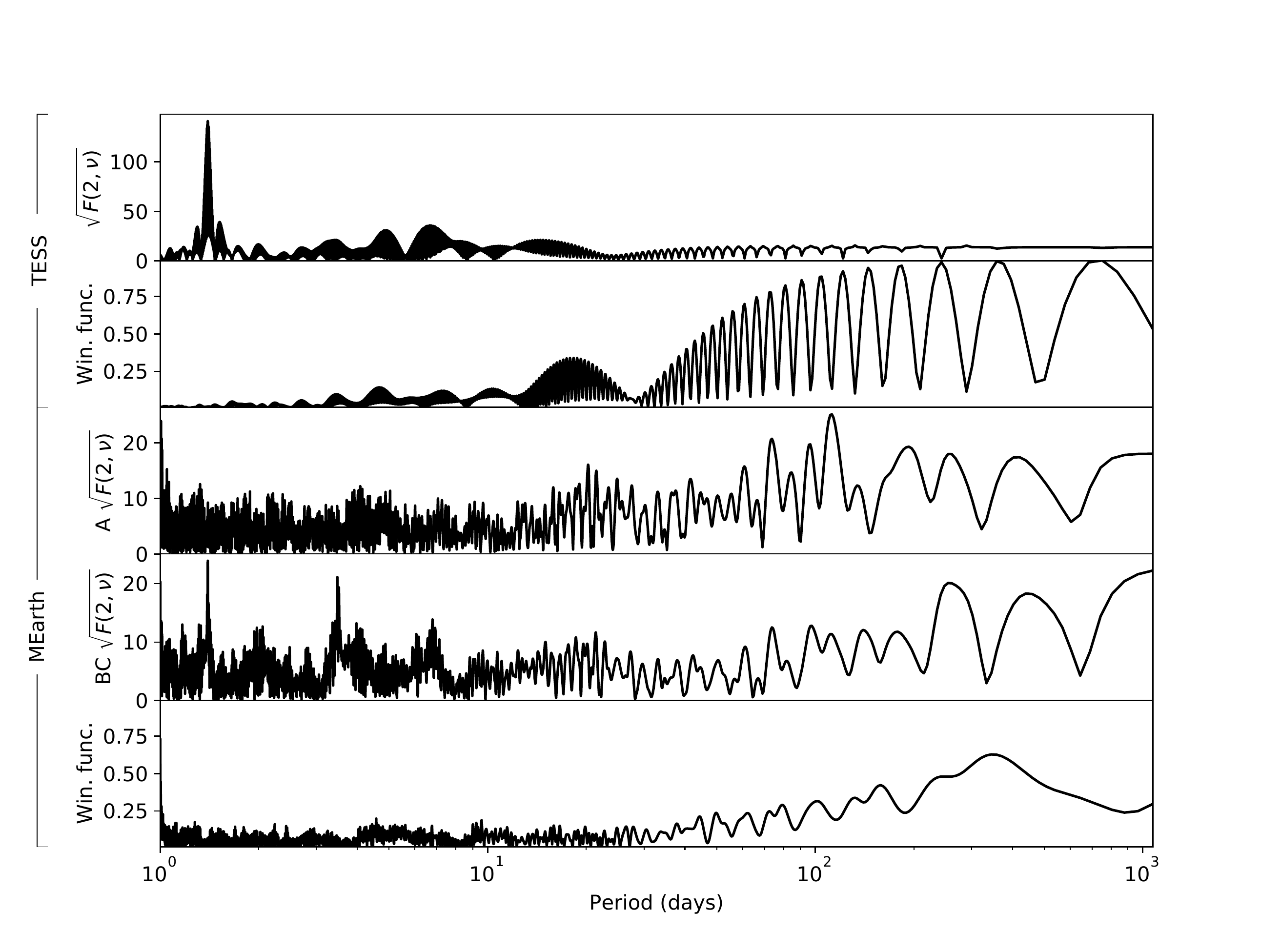}
\caption{Periodograms and Window Functions of \tess ~and MEarth data of LTT~1445. The \tess ~data (top two panels), which include all three stellar components, exhibit a 1.4-day period. This signal is also seen in the periodogram of the MEarth data for LTT~1445BC (fourth panel), confirming that either the B or C component is the origin of the 1.4-day signal in the TESS data. The peak at 3.5 days represents an alias of the 1.4-day period. The cluster of peaks between 50 and 100 days likely indicate the rotation period of A (middle two panels), but more data are needed for a robust detection of the expected period of $85 \pm 22$ days. A marginal 6.7-day signal is also seen in both the \tess ~data and the MEarth data of LTT~1445BC, which we discuss further in \S \ref{subsec:bc_rot_per} . \label{fig:pgrams}}
\end{figure*}

The resulting periodograms are shown in Figure \ref{fig:pgrams}, where we compare
results for the TESS light curve containing all three components with
the separate MEarth light curves for A and BC. We also show the
periodogram of the observational window function in each case. 

The periodogram is plotted in terms of the F-test statistic for the
nested model comparison between the null hypothesis of only the
baseline detrending with no rotation signal, and the alternate
hypothesis where there is also a sinusoidal signal of the given period
(further details of this procedure are given in \citealt{Newton(2016)}). At
fixed period, this statistic would follow an $F$ distribution $F(2,
\nu)$ where the number of degrees of freedom $\nu$ is large. Standard
results for this distribution show the 1.4 day signal in the MEarth
light curve of LTT 1445 BC is highly significant with an $F$ value
several hundred above the baseline level in the periodogram.

\begin{figure}
\centering
\includegraphics[scale=.32,angle=0]{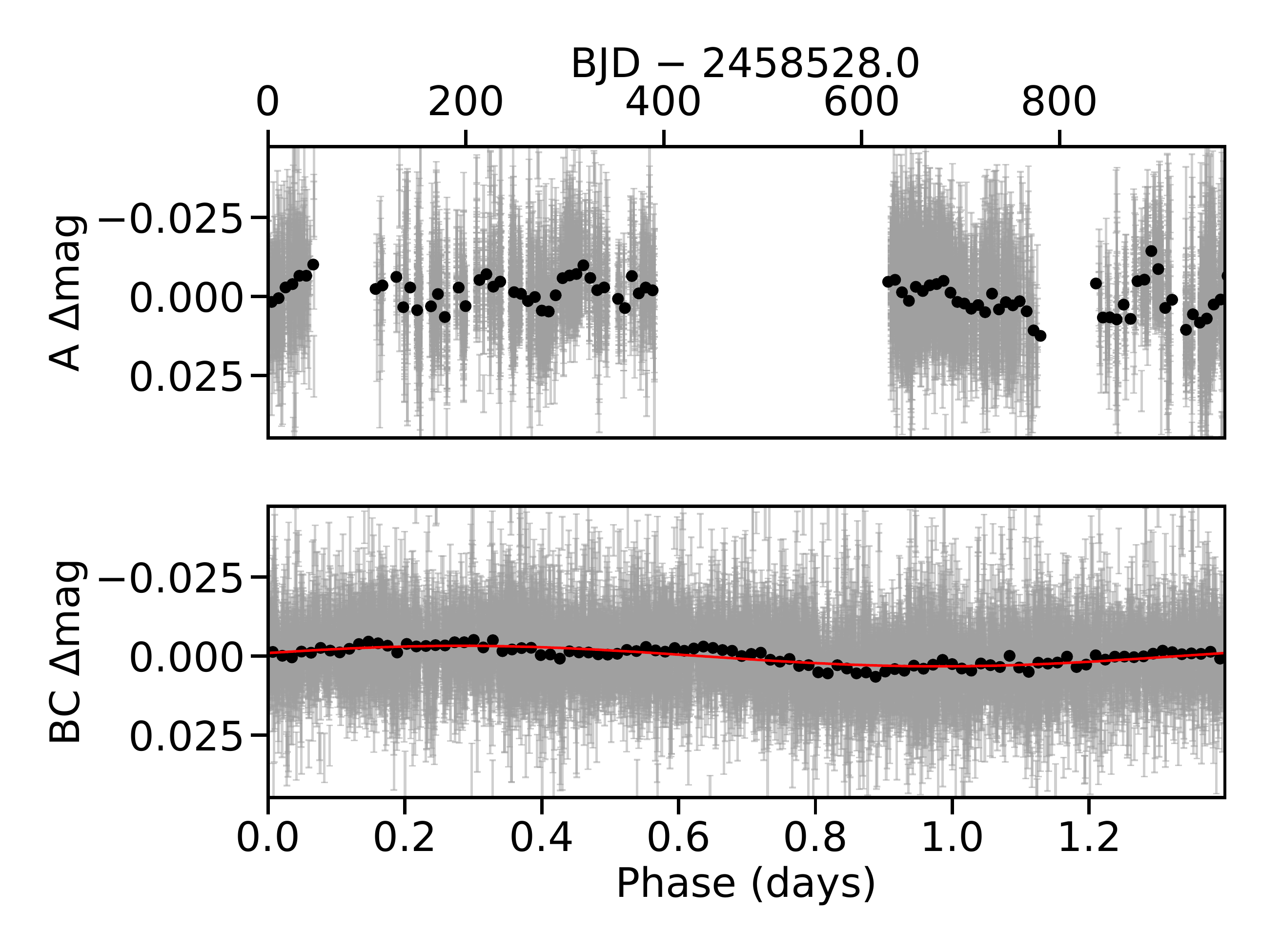}
\caption{MEarth light curves of LTT~1445A (top) and BC (bottom), phase-folded to the 1.4-day stellar rotation period. The data have been corrected for precipitable water vapor, and outliers have been rejected. The gray points and errorbars indicate the individual measurements, while the black points indicated the binned data. The red curve in the bottom panel illustrates the 1.4-day rotation period model for LTT~1445BC. The top panel illustrates that we do not yet have enough data for a robust rotation period detection of LTT~1445A.   \label{fig:ltt01445_mearth_prots}}
\end{figure}

The 1.4-day signal detected in TESS data and reported previously by
\citetalias{Winters(2019b)} is found to originate from BC, as expected
based on stellar activity and rotational broadening results presented
in the previous paper, but this has now been resolved observationally.
The $1\ {\rm day}^{-1}$ alias of this signal is also seen close to 3.5
days.  No clear rotational modulation signal meeting our quality
criteria (see \citealt{Newton(2016)}) is yet detected in LTT
1445A, but the periodogram shows power at long periods between
approximately 40 and 150 days and the corresponding alias period close
to 1 day.  Based on the observed activity level and mass of this star,
a rotation period around $85\pm22$ days would be
expected (calculated using the empirical relation in \citealt{Newton(2017)}), but has yet to be detected in the observations. We show the MEarth light curves for LTT~1445A and BC in Figure \ref{fig:ltt01445_mearth_prots}. 

\subsubsection{MEarth-South Transit Follow-Up Observations}

We also performed follow-up observations with MEarth-South spanning times of the transit of planet b. Because MEarth is able to spatially resolve star A from stars BC, these data serve to confirm that planet b indeed transits star A. As discussed above in \S \ref{subsubsec:mearth_south_photometric_monitoring}, MEarth
observations of the LTT~1445 system are extremely sensitive to image
quality due to the components being barely resolved in the data.
Additionally, due to the short exposure time, atmospheric
scintillation noise is severe and increases with airmass.
Consequently, although we observed several transits of LTT~1445Ab for
followup, the majority of these were of poor quality, and all but one
were rejected for the final analysis.

Our analysis includes a single full transit observed on UT 2019
September 2 using 7 telescopes.  Otherwise, the observational strategy
and reduction techniques were identical to those described in
\citetalias{Winters(2019b)} and are not repeated here. A total of
3935 data points were gathered during this time series and the
resulting light curve is shown in the top panel of Figure \ref{fig:ground_transits}.

Because the expected transit depth of the 3.12-day planet candidate was below the detection limits of the MEarth array, we did not attempt observations to confirm it with MEarth.

\subsection{LCOGT Transit Follow-up Observations}
\begin{figure}
\centering
\includegraphics[scale=.52,angle=0]{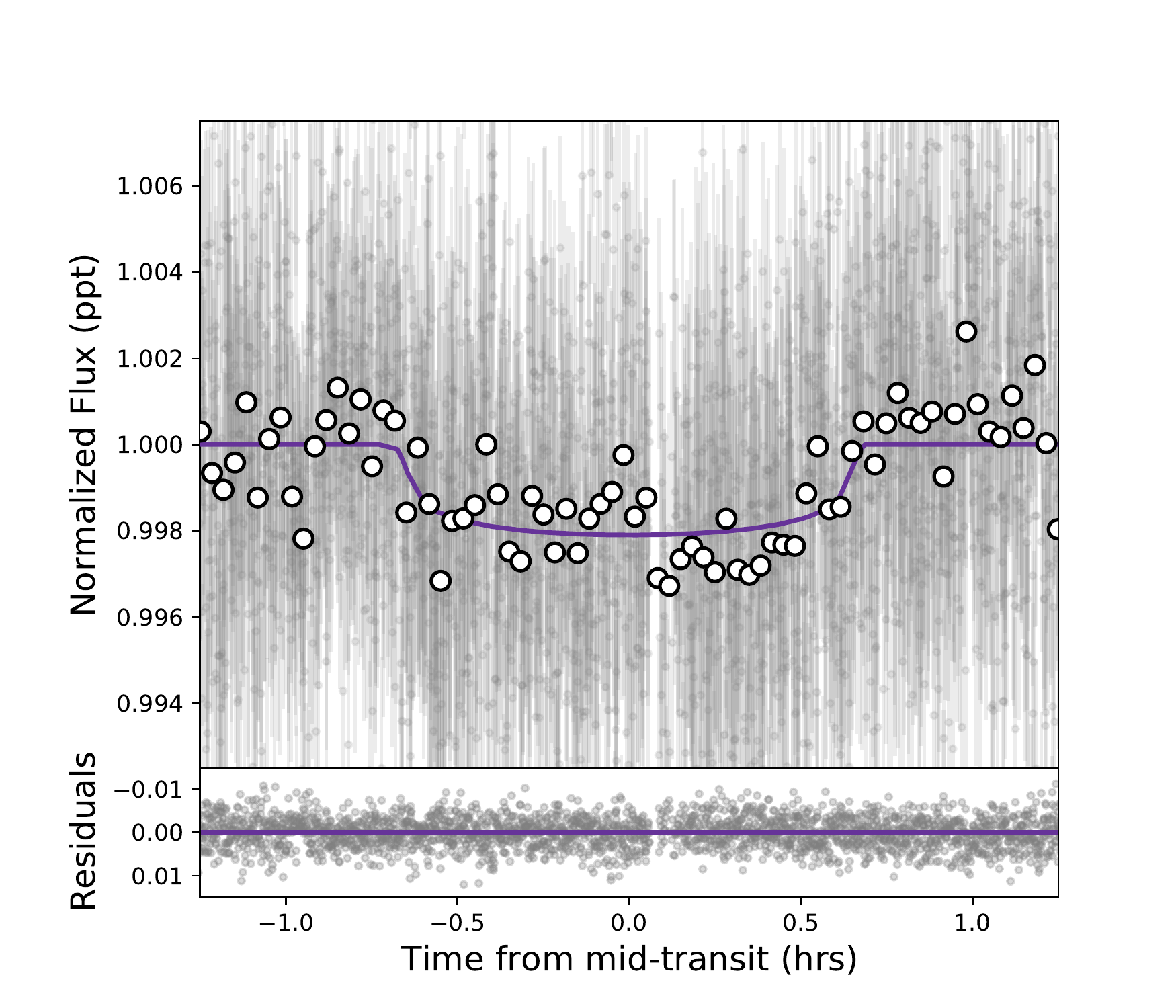}
\includegraphics[scale=.30,angle=0]{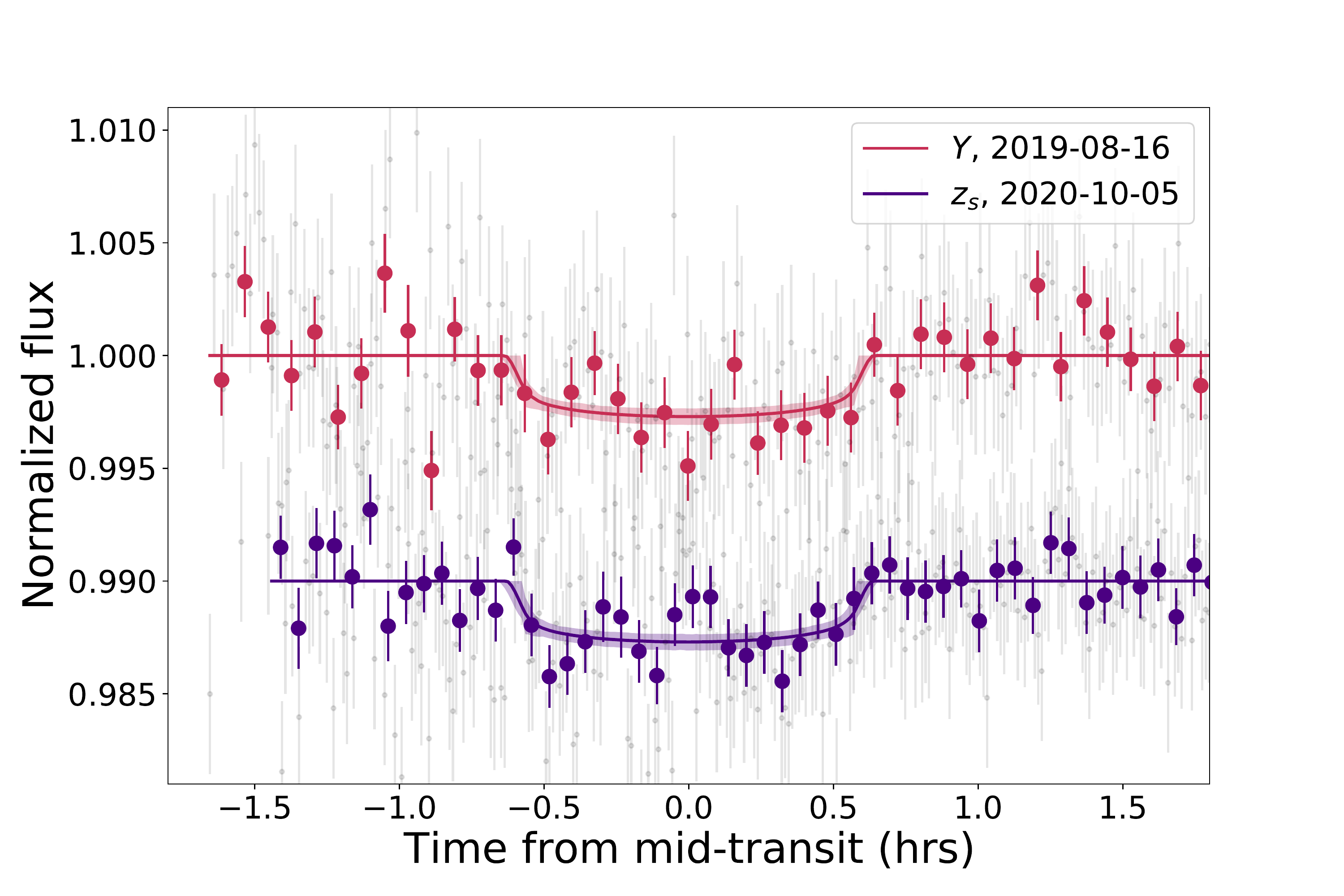}
\caption{Transits of LTT~1445Ab measured from ground-based data. The top panel shows the transit of LTT~1445Ab as observed by MEarth-South on UT 2019 September 2 using 7 telescopes. The bottom panel shows two transits of LTT~1445Ab as observed by LCOGT. \label{fig:ground_transits}}
\end{figure}

We observed three full transits of LTT~1445Ab from the Las Cumbres Observatory Global Telescope (LCOGT) \citep{Brown:2013} 1.0\,m network. We observed a full transit on UT 2019 August 16 in Pan-STARRS Y-band from the CTIO node using 30 second exposures, and a full transit on UT 2020 October 8 in Pan-STARRS $z$-short band from the McDonald Observatory node using 10 second exposures. A full transit was observed on UT 2019 July 16 from the South Africa Astronomical Observatory node in Pan-STARRS Y-band using 30 second exposures. However, we ultimately chose not to include this observation in the analysis because only a few minutes of nominal post-transit baseline were observed, resulting in an unclear time of egress and thus a depth measurement that is not well constrained. We used the {\tt TESS Transit Finder}, a customized version of the {\tt Tapir} software package \citep{Jensen:2013}, to schedule our transit observations. The $4096\times4096$ LCOGT SINISTRO cameras have an image scale of $0\farcs389$ per pixel, resulting in a $26\arcmin\times26\arcmin$ field of view. The images were calibrated by the standard LCOGT {\tt BANZAI} pipeline \citep{McCully:2018}, and photometric data were extracted with {\tt AstroImageJ} \citep{Collins:2017}. The UT 2019 August 16 images were (unintentionally) mildly defocused and have typical stellar point-spread-functions with a FWHM of roughly $3\farcs 8$. Circular apertures with radius of roughly $2\farcs 7$ were used to extract the differential photometry. We estimate that about $2\%$ of the flux in the target star aperture is from the nearby pair LTT~1445BC. The UT 2020 October 8 images were very slightly defocused and have typical stellar point-spread-functions with a FWHM of roughly $2\farcs 1$. Circular apertures with radius about $3\farcs 5$ were used to extract the differential photometry. We estimate that about $1.3\%$ of the flux in the target star aperture is from the nearby pair LTT~1445BC. For all observations, we used a single comparison aperture around the nearby pair LTT~1445BC to extract the light curve of LTT~1445Ab. The light curves are presented in the bottom panel of Figure \ref{fig:ground_transits}.

Using the LCOGT 1.0\,m network and an ephemeris extracted from $TESS$ sector 4 data, we attempted to recover the 3.12-d transit signal on 6 epochs in 2019 and 2020. However, the updated ephemeris based upon the combination of the sector 4 and 31 data shows that those observations were typically 5 hours late and do not span the times of transit. 

\subsection{Spectroscopic Data}

We combined 136 RVs of LTT~1445A from five high-resolution spectrographs taken over roughly two years. Here we describe the instruments in alphabetical order, as well as the methods used to measure the RVs and their uncertainties. We list the RVs and their uncertainties in Table \ref{tab:rvs}. 
 
\subsubsection{ESPRESSO}

We acquired 19 spectra of the LTT~1445A with the Echelle SPectrograph for Rocky Exoplanets and Stable Spectroscopic Observations (ESPRESSO) spectrograph \citep{Pepe(2021)} on the VLT from UT 2020 July 04 to September 03. ESPRESSO is an ultra-stable, fiber-fed echelle spectrograph with resolving power $R=$ 140,000 in the high-resolution (1-UT) configuration and a wavelength range of $378.2 - 788.7$ nm. We used integration times of 900 sec and we employed the slow (100 kpx \pers), $2\times1$ binning readout mode. We elected not to use the simultaneous wavelength calibration; thus, the calibration fiber was set on sky. The signal-to-noise ratios (SNR) of the spectra range between 58 and 142 at 650 nm, with an average of 104. 

We derived the RVs using the $\chi^2$-minimization template-matching method described in \citet{Astudillo-Defru(2015)}. The stellar template was built from the median of the individual spectra, accounting for the ESPRESSO design that creates
two individual images (slices) of the same spectral order. To compute the RVs, we discarded echelle orders below order 15 (30 slices) due to the low flux, as well as zones containing telluric features. RV uncertainties were computed following the methods described in \citet{Bouchy(2001)} and range between 0.15 m \pers ~and 0.32 m \pers, with an average uncertainty of 0.20 m \pers.

\subsubsection{HARPS}
\label{subsubsec:harps}

We acquired 45 spectra of LTT~1445A with the High Accuracy Radial Velocity Planet Searcher (HARPS) spectrograph \citep{Mayor(2003)} on the La Silla 3.6-m telescope from UT 2019 March 04 to October 21. HARPS is a fiber-fed echelle spectrograph with resolving power $R=$ 115,000 and a wavelength range of 378-691 nm. We employed integrations times of 900 sec and the spectra were read out in slow mode (104 kpx \pers), except for one spectrum acquired on 2019 September 12, which had an exposure time of 1500 sec. We found that for seven of our spectra we had inadvertently observed the BC component, as evidenced by the H$\alpha$ emission. We discarded these from the RV analysis, but these spectra later proved to be a blessing in disguise, as they allowed us to investigate the rotational broadening of the B and C components (see \S \ref{subsec:bc_rot_per}). The calibration fiber was set on sky. The science fiber provides SNRs between 26 and 60 at 650 nm, with an average of 48. As with ESPRESSO, we derived the RVs by template matching and we estimated the uncertainties following the methods of \citet{Bouchy(2001)}; these range between 1.06 m \pers ~and 2.86 m \pers, with an average uncertainty of 1.44 m \pers. We allowed for a separate RV zero point offset for the 2019 February data because of the roughly four month time span between those data and the later observations. 

We also measure the Ca II H and K emission (S activity index) and proceed to calculate a value of log(R'$_{HK}$)=$-5.413\pm0.118$. Following \citet{Astudillo-Defru(2017)}, we estimate a rotation period of $79\pm15$ days, in agreement with the estimated rotation period of $85\pm22$ days noted above. 

\subsubsection{HIRES}

We acquired 39 spectra of LTT~1445A with the High Resolution Echelle Spectrometer (HIRES) on the 10-m Keck telescope on Maunakea, Hawaii, on nights between UT 2019 August 14 and 2020 February 28. HIRES is an optical echelle spectrograph with resolving power $R \approx 55,000$ \citep{Vogt1994}.  We used the standard procedures of the California Planet Search \citep[CPS;][]{Howard2010} to make the observations and reduce the spectra.  This includes the use of an iodine cell to calibrate the wavelength solution and point spread function of HIRES, as well as a long slit (the ``C2'' decker) to measure and subtract contemporaneous sky spectra.  We used an image rotator to orient the field so that light from LTT~1445A went through the slit and light from LTT~1445BC did not. 
The HIRES spectra span the wavelengths 364--790 nm with our CPS setup, but the RVs were computed from the wavelength range 500--630 nm because that region is calibrated by the iodine. 
We used integration times of 900~s, which resulted in SNRs of 100--170 per pixel on blaze near 550 nm.  
We computed relative RVs using the method descended from \cite{butler1996}.  The HIRES RVs have internal uncertainties between 1.3 m \pers ~and 2.0 m \pers, with an average uncertainty of 1.5 m \pers.

\subsubsection{MAROON-X}
We used the MAROON-X spectrograph \citep{Seifahrt(2016), Seifahrt(2018), Seifahrt(2020)} on the 8.1-m Gemini-North telescope to measure RVs of LTT~1445A during three observing runs of two weeks each between December 2019 and November 2020. MAROON-X is a stabilized, fiber-fed echelle spectrograph with resolving power $R \simeq$ 85,000 and a wavelength range of $500 - 920$\,nm in two camera arms. MAROON-X has demonstrated an RV stability of at least 30\,cm\,s$^{-1}$ over the span of a few weeks during its first year of operations \citep{Seifahrt(2020)} and has recently been used to determine a precise mass of the nearby transiting rocky planet Gl\,486b \citep{trifonov21}.

We acquired 27 spectra of LTT~1445A over 8 nights during the science verification and commissioning phase of the instrument in December 2019. The simultaneous calibration was not operational during these first observations, so bracketing wavelength calibration exposures were taken to track the instrumental drift. The lack of simultaneous calibration likely limits the instrumental precision of these data to approximately 1\,m\,\pers. Moreover, we found that 7 of the 27 spectra in two consecutive nights in 2019 had low SNR and also showed abnormal chromatic index and differential line widths in MAROON-X's blue arm. Some of these spectra lead to RV outliers and thus all 7 spectra were subsequently dropped from the RV analysis. Due to their low SNR and thus larger individual RV uncertainty, their contribution to the nightly mean was small and did not significantly change the final RV data point for the night.

We gathered an additional 20 spectra over 17 nights in September and November 2020. We find only a single night with notable stellar activity in this dataset. Because the RV measurement from this one night does not present as a clear outlier, we included this datapoint in the RV analysis.

We used integration times of 600\,sec in 2019 and 300\,sec in 2020. In 2019, the peak SNR per 1D pixel ranges between 60 and 190 at 640\,nm in the blue arm and between 140 and 430 at 800\,nm in the red arm, with a widespread distribution depending on seeing conditions. In 2020, we find peak SNRs ranging between 50 and 95 at 640\,nm in the blue arm and between 130 and 250 at 800\,nm in the red arm.

We reduced the MAROON-X raw data using a custom \texttt{Python~3} pipeline based on tools previously developed for the CRIRES instrument \citep[][]{Bean(2010)}. The wavelength and instrumental drift solution of MAROON-X is based on a stabilized Fabry Perot etalon  \citep[][]{Stuermer(2017)} in conjunction with a ThAr lamp for the initial measurement of the gap size and chromatic dispersion of the etalon. In early 2020, the simultaneous calibration fiber of MAROON-X became operational, allowing for a robust order-by-order drift correction at the sub-m \pers\ level. Barycentric corrections were computed using the \texttt{barycorrpy} code \citep[][]{Kanodia_2018} from weighted averages based on MAROON-X's white light exposure meter.

We computed the RVs using \texttt{serval} \citep[][]{Zechmeister(2018)}, which employs a $\chi^2$-minimization algorithm based on matching an empirical template to each order at each epoch. The template is constructed from the data itself as a weighted average of all observations. A telluric mask is used to exclude telluric absorption features deeper than about 1\%, both in the construction of the template and in each epoch. Typical uncertainties derived by \texttt{serval} are based on the inter-order dispersion of the RVs and range from 0.25\,m \pers\ for the highest SNR of 430 in the red channel to 1.6\,m \pers\ for the lowest SNR of 60 in the blue channel. To increase the SNR for robust RV calculations, we combined blue and red arm observations as well as multiple back-to-back observations per night into nightly epoch means and propagated uncertainties based on the spread of the individual measurements. These final uncertainties range from 0.4 to 1.4\,m \pers\ in 2019 and 0.4 to 1.0\,m \pers\ in 2020, with an average uncertainty of 0.8 and 0.6 m \pers, respectively. 

Because of an instrument intervention that took place after the 2019 commissioning run, which introduced changes to both the etalon as well as the spectral format of the blue arm, we treat the data from 2019 and 2020 as independent data sets with distinct RV offset parameters. 

\subsubsection{PFS}

We acquired 25 spectra (15 when binned by night) of LTT~1445A with the Planet Finder Spectrograph \citep[PFS,][]{crane2006,crane2008,crane2010}, mounted on the 6.5-m Magellan II telescope at Las Campanas Observatory in Chile, from UT 2019 July 20 to 2020 November 4. PFS covers $391 - 734$ nm and the subsequent RVs are calibrated using the iodine method. The default 0\farcs3 slit provides a resolving power $R\simeq 127,000$. Integration times ranged from 960 to 1500 s resulting in SNRs between 40 and 70 at the peak of the blaze around 600 nm. All PFS spectra are reduced and analyzed using a custom {\texttt {IDL}} pipeline based on \citet{butler1996} that regularly delivers sub-1 m~s$^{-1}$ precision. We binned observations taken sequentially to improve the SNR for robust RV calculations. The RV uncertainties in our nightly-binned RVs ranged between 0.48 m \pers ~and 0.83 m \pers, with an average uncertainty of 0.63 m \pers. 

\begin{deluxetable}{lccc}
\tabletypesize{\small}
\tablecaption{Radial Velocities for LTT~1445A  \label{tab:rvs}}
\tablecolumns{4}
\tablenum{1}

\tablehead{
\colhead{BJD\tablenotemark{a}} & 
\colhead{$v_{\rm rad}$\tablenotemark{b}} &
\colhead{$\sigma$} &
\colhead{Spectrograph}  \\ 
\colhead{(days)} & 
\colhead{(${\rm m\ s^{-1}}$)} &
\colhead{(${\rm m\ s^{-1}}$)} &
\colhead{}
}
\startdata
2458546.500296  & -5457.74  &    1.31  & HARPS \\
2458547.501338  & -5463.53  &    1.16  & HARPS \\
2458548.500633  & -5460.85  &    1.36  & HARPS \\
2458555.506869  & -5455.18  &    1.29  & HARPS \\
2458556.505906  & -5453.54  &    1.16  & HARPS \\
\enddata
\tablenotetext{a}{Barycentric Julian Date of mid-exposure, in the TDB time-system.}
\tablenotetext{b}{Barycentric radial velocity.}
\tablecomments{This table is available in its entirety in machine-readable form.}

\end{deluxetable}

\section{Analysis}
\label{sec:analysis}

Because of the complex nature of the \tess ~light curve, we used a combination of {\texttt{exoplanet}} \citep{Foreman-Mackey(2017)} and {\texttt{ExoFASTv2}} \citep{Eastman(2019)} for our light curve modeling: We first used {\texttt {exoplanet}} to fit and remove the photometric modulation in the \tess ~light curve using a Gaussian Processes (GP) regression model via {\texttt {celerite}} while preserving the transit signals of the two planets. We then used {\texttt{ExoFASTv2}}, which does not use GPs, to simultaneously fit the RV data and the de-trended transit data from {\texttt {exoplanet}}. We also performed an independent RV-only analysis that included a GP regression to account for stellar activity that could affect the uncertainties of the planetary masses.

We will know the masses and radii of these planets only as well as we know these values for their host star. We estimate a mass of $0.257\pm0.014$ M$_{\odot}$ for the host star using the mass-luminosity relation in the $K-$band by \citet{Benedict(2016)} and then use a single-star mass-radius relation \citep{Boyajian(2012)} to estimate a stellar radius of $0.268\pm0.027$ R$_{\odot}$. We calculate the bolometric corrections in $K$ \citep[erratum]{Mann(2015)} and in $V$ \citep{Pecaut(2013)}, along with their respective bolometric luminosities. We adopt the mean of the two bolometric luminosities and then use the Stefan-Boltzmann Law to calculate an effective temperature $T_{\rm eff}$ of $3337\pm150$ K for LTT~1445A. We adopt the metallicity [Fe/H] of $-0.340\pm0.090$ dex from \citet{Neves(2014)}, which is measured using HARPS data. We updated the parallax of LTT~1445A to that of the Gaia EDR3 \citep{Gaia(2016a), Lindegren(2021)} for our estimates of the star's mass, radius, and effective temperature, but these values did not change significantly from those reported in \citetalias{Winters(2019b)}. 
We use these stellar parameters of LTT~1445A as priors for the light curve and RV modeling described below.

\subsection{Light Curve Modeling with \texttt{exoplanet}}

The presence of other objects in the \tess ~aperture can dilute the transit depth of the planet, which is measured from the light curve. If the additional objects have identifiers and \tess ~magnitudes in the \tess ~Input Catalog \citep[TICv8.1;][]{Stassun(2019)}, the SPOC pipeline calculates a dilution (i.e., contamination) factor and performs a correction to the final light curve \citep{Stumpe(2012)}. For the sector 4 data, the SPOC contamination value of 0.4849 is very close to our estimated value of $0.480 \pm 0.013$, which is calculated from $T$ mags estimated from an M-dwarf-specific relation \citepalias{Winters(2019b)}. For the sector 31 light curve, the SPOC contamination value is 0.42003, which is significantly different. The difference is partly due to an updated $T$ mag for LTT~1445A of $T = 8.843$ mag in the TICv8.1, compared to $T = 8.64$ mag from an earlier TIC version. However, the TICv8.1 also appears to have a duplicate entry for LTT~1445BC, one from the 2MASS cross-match (TIC 98796342) with $T = 8.554$ mag and the other from the Gaia DR2 cross-match (TIC 651819442) with $T = 11.558$ mag. Thus, the SPOC dilution value in the sector 31 light curve is an overestimate of the amount of contaminating light from the BC components. As in \citetalias{Winters(2019b)}, we removed the SPOC dilution value and imposed our own.  

As in \citetalias{Winters(2019b)}, we did not include the third transit of LTT~1445Ab in sector 4 in our analysis because the light curve baseline showed a strong slope at egress; we note that this transit was also omitted from the results in the SPOC DV report. The third transit of LTT~1445Ac in the sector 4 light curve was not observed, as it occurred during the communication failure. The fourth transit occurred during the brief chunk of light curve right before the data download. Our GP did not fit properly over such a small span of data, so in the interest of working with uniformly processed data, we omitted this transit from our analysis. The third transit of the 5.36-day planet in the sector 31 data occurred during the pause in data collection while the data were being downloaded and was thus not observed. The star tracker anomaly reported for sector 31 did not affect the transits of either planet.  

We used the python package {\texttt {exoplanet}} \citep{exoplanet:exoplanet}, which employs probabilistic methods to model exoplanet transit and radial velocity data sets. It has the additional capability to incorporate Gaussian Processes (GP) with {\texttt {celerite}} \citep{Foreman-Mackey(2017)} and limb-darkened light curves with {\texttt {starry}} \citep{Luger(2018)}.  We used the SPOC-generated Pre-Search Data Conditioning Simple Aperture Photometry (PDCSAP) light curve \citep{Smith(2012),Stumpe(2012),Stumpe(2014)}, corrected with our calculated dilution factor. Before fitting, we removed positive outliers (flares) that deviated by more than three times the median absolute deviation of the PDCSAP light curve. While we do not know from which star the flares originate, recent work on nearby mid-to-late M dwarfs (see, e.g., \citealt{Medina(2020)}) indicates that rapidly rotating stars flare far more frequently. Thus, we suspect that many of the flares originate from either or both of the B or C components. 

We model the stellar rotational modulation, as well as any other possible systematics, with a GP. The GP kernel is the sum of two simple harmonic oscillators, which has been shown to be an appropriate kernel for data that are quasi-periodic in nature \citep{Angus(2018)}, such as the observed rotational modulation in the light curve of LTT~1445ABC. To model the planetary transits, we used a limb-darkened transit model and a Keplerian orbit. We used the same model, priors, GP hyper-parameters, and methods for the fit as described in \citetalias{Winters(2019b)}. We measure planet-to-star radius ratios of $0.037\pm0.005$ and $0.047\pm0.006$ for LTT~1445Ac and LTT~1445Ab in the $TESS$ sector 4 light curve data and ratios of $0.035\pm0.009$ and $0.042\pm0.007$ for the respective planets in the $TESS$ sector 31 light curve data. 

\begin{figure*}
\centering
\includegraphics[scale=.70,angle=0]{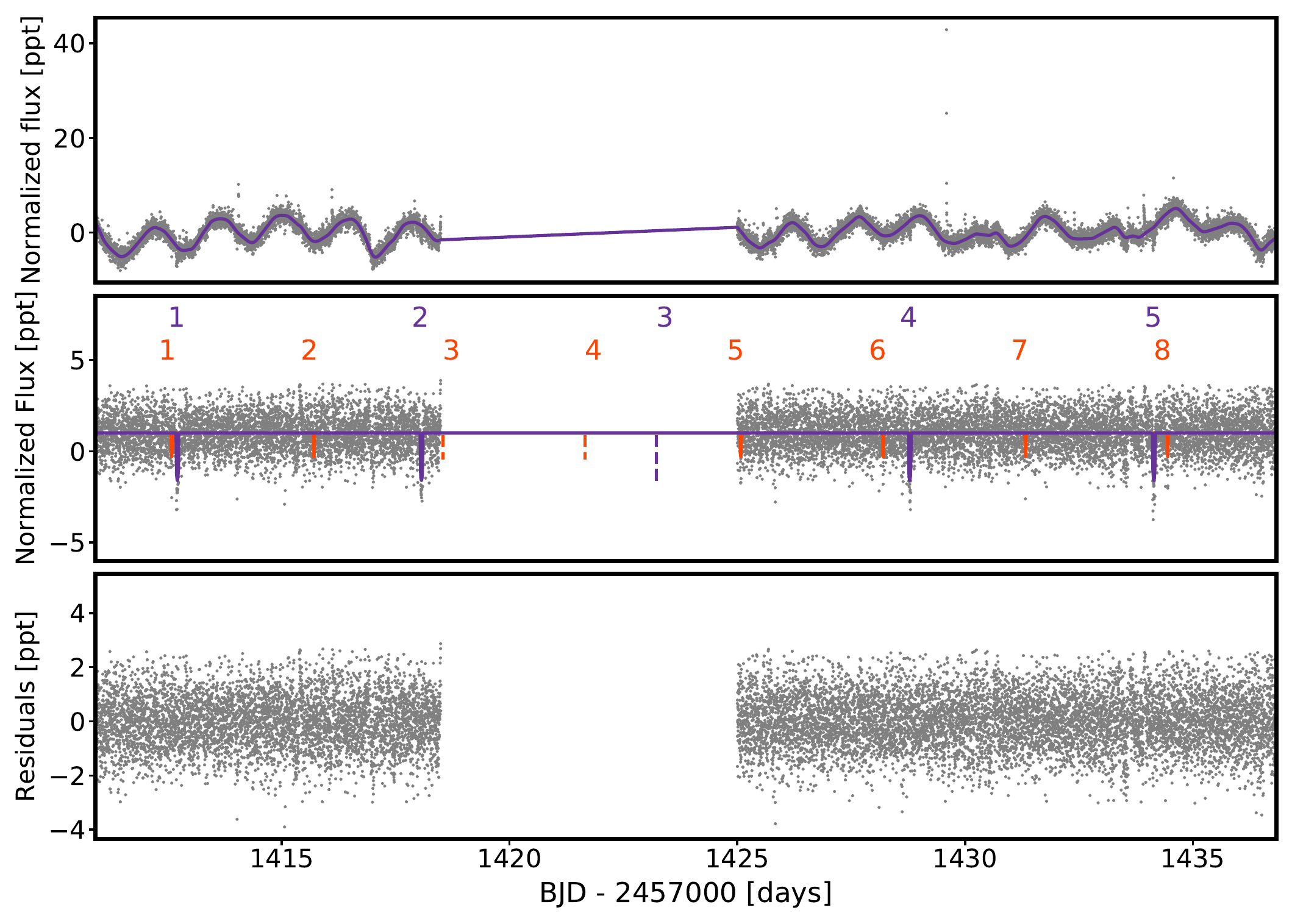}
\includegraphics[scale=.70,angle=0]{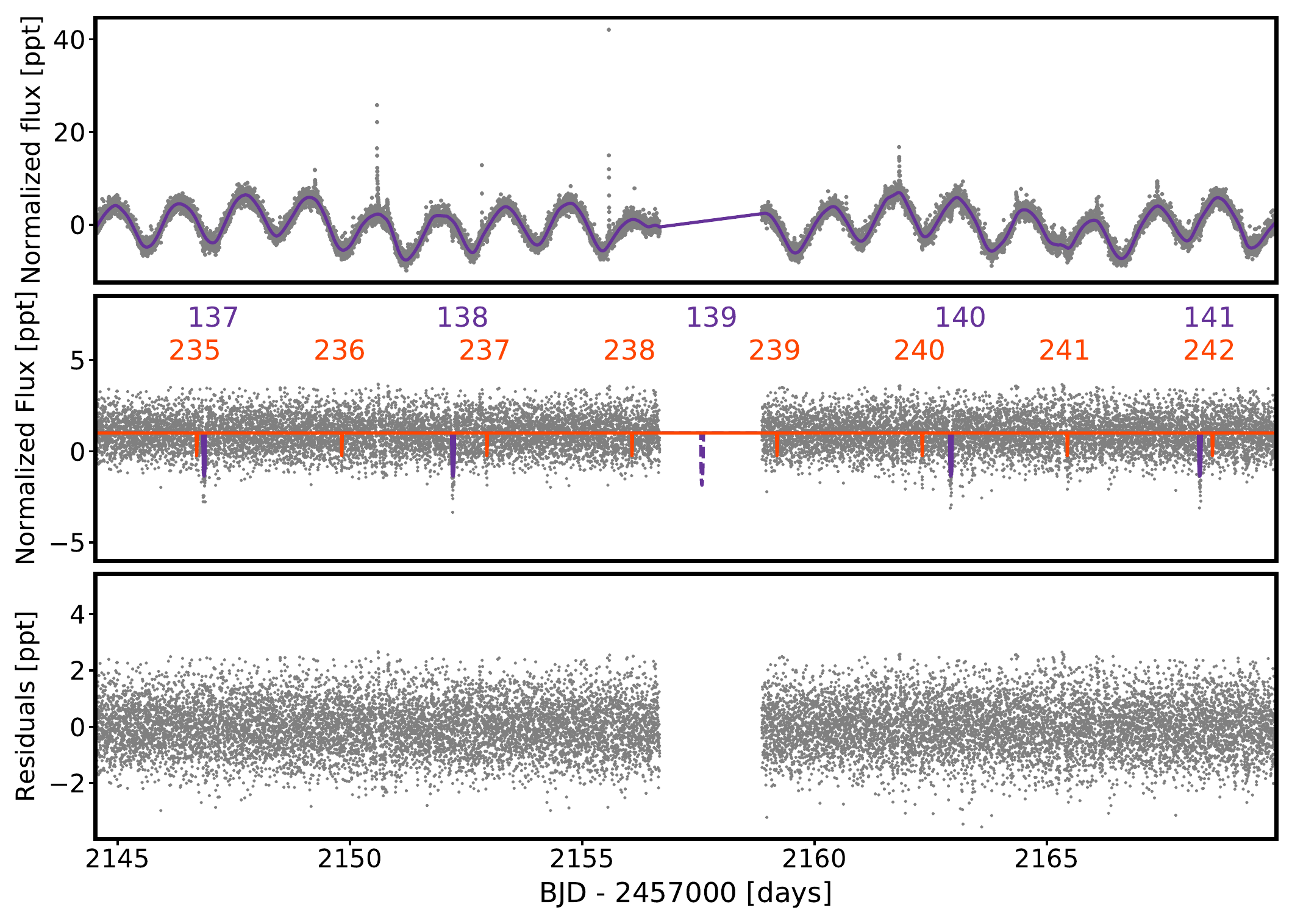}
\caption{The \tess ~PDCSAP light curves from sector 4 (top 3 panels) and 31 (bottom 3 panels) data. The top panels of each plot are the non-detrended light curves from each sector of data, which show evidence of flares and rotational modulation due to stellar spots on either the B or C component. The purple lines indicates the fits to the modulations. The middle panels of each plot show the residual data after the removal of the stellar variability with Gaussian Process (GP) regression; the planetary transit models for the 5.36-day planet (purple line) and 3.12-day planet (orange line) are overplotted. The dotted lines indicate transits that occurred during a data download, and which were not observed. The bottom panels of each plot illustrate the light curves with planetary transits and stellar variability removed.   \label{fig:medina_gp}}
\end{figure*}

We note that the 1.4-day rotation signal has evolved in the two years that elapsed between sector 4 and 31. While the period of the modulation is the same, the semi-amplitude of the variability has roughly doubled from $1.475\pm0.302$ ppt in the sector 4 light curve to $3.027\pm0.401$ ppt in the sector 31 light curve. This increase in the semi-amplitude is likely due to a change in the morphology of the starspot groups leading to a larger hemisphere-integrated asymmetry in sector 31. We show our GP fit to the \tess ~light curve data for both sector 4 and 31 in Figure \ref{fig:medina_gp}.

\subsection{Global Modeling with \texttt{ExoFASTv2}} \label{subsec:global}

We perform a simultaneous global fit of the \tess ~light curve and radial velocity data using {\texttt{ExoFASTv2}} \citep{Eastman(2019)}. {\texttt{ExoFASTv2}} is a suite of {\texttt {IDL}} routines that simultaneously fits exoplanetary transit and radial velocity data using a differential Markov Chain Monte Carlo (MCMC) code. We include the \tess ~light curves from sectors 4 and 31, and we use 134 RVs from the five spectrographs. We used as input for the {\texttt{ExoFASTv2}} the output light curve data from {\texttt {exoplanet}} with the stellar variability removed. In order to allow for the propagation of the uncertainty in the dilution that was corrected by the {\texttt {exoplanet}} fit, we allow for a residual dilution, with a prior centered on zero with $\sigma = 0.013$. 

Because we derived the stellar parameters as described in \S \ref{sec:host_system}, we did not include a spectral energy distribution in the fit, and we disabled the default MIST stellar evolutionary models that use isochrones to constrain the stellar parameters. We placed Gaussian priors on the stellar mass, radius, effective temperature, and metallicity that were equal to the uncertainties noted above in \S \ref{sec:analysis}. We interpolated the quadratic limb darkening coefficients in Table 15 of \citet{Claret(2017)} to the effective temperature of LTT~1445A, which we imposed with a Gaussian prior with a standard deviation of 0.10. While the atmospheric models used to derive the limb darkening tables are questionable for low-mass stars such as \star, the impact is likely to be negligible due to the low precision of the \tess ~light curve. 

We include a slope term to account for a change in the RVs of LTT~1445A due to the stellar A-BC orbit. We estimated the magnitude of the RV slope as follows. Because the maximum change in RVs over the 643-day timespan of our data is dependent upon the eccentricity of the orbit, we explored which eccentricities would result in a stable configuration for the A-BC orbit. The stability criterion for hierarchical triple star systems from \citet{Harrington(1972)} requires the eccentricity of the outer (A-BC) orbit and the ratio of the semi-major axes of the inner (B-C) and outer orbits. We investigated eccentricites 0 to 0.9 and semi-major axes of 1\arcsec ~(from the preliminary astrometric orbit for LTT~1445BC reported in \citet{Winters(2019b)}) and 5\arcsec ~(the mean separation noted for LTT~1445A-BC in \S \ref{sec:host_system}), respectively, to estimate that retrograde motion orbits with eccentricities less than 0.45 would be stable. Therefore, we estimate that a orbital period of 253 yr for LTT~1445A-BC would result in a drift in our RV data with values between 0.0002 m \pers ~day$^{-1}$ for a circular orbit to 0.25 m \pers ~day$^{-1}$ for an eccentric ($e = 0.45$) orbit. We ran two separate {\texttt {ExoFASTv2}} fits, one that included the term for the slope, and one that did not. {\texttt{ExoFASTv2}} provides statistical metrics, such as the Bayesian Information Criterion (BIC) \citep{Schwarz(1978)} and Akaike Information Criterion (AIC) \citep{Akaike(1974)}, for model fits. Both metrics are functions of the number of parameters being fit and the maximum ln(likelihood) from the MCMC sample, while the BIC also considers the number of data points being fit. The better model has the smallest algebraic value of AIC or BIC. The fit that included the slope term for the RVs was slightly preferred over the flat RV fit, with BIC and AIC values of $-22202$ and $-22424$, respectively, for the model with no slope, compared to values of $-22209$ and $-22437$ for the model with the slope. We thus included the RV slope term in all of our models.

We rejected flat models (i.e., any instances where an expected transit was not found) in order to prevent the fit from running away, which can happen in the hotter chains when using parallel tempering. We required the number of independent draws to be greater than 1000 and determined that, with a Gelman-Rubin statistic \citep{Gelman(1992)} of 1.0090 in the worst case, the chains were well-mixed. 

We initially allowed the radius of the 3.12-day planet to be unconstrained, but found that grazing geometries were permitted, indicating that the light curve can set only a lower limit on the planet's radius and an upper limit on its inclination. Similar to \citet{Rodriguez2018}, we imposed a constraint on the radius of the 3.12-day planet based on the mass-radius relations by \citet{Chen(2017)}. The impact is that this prior sets an upper limit on the planet radius (resulting from the measured planetary mass, informed by the RVs), while the lower limit on the planet radius is set from the observed depth of the transit. We emphasize that our data do not permit us to directly determine the radius of the 3.12-day planet: Although we state the value we determine by employing the \citet{Chen(2017)} prior, the planet radius could be much larger and should be determined by gathering data of greater precision, which would permit the determination of the four points of contact (if they exist). In Table \ref{tab:master_table}, we have flagged all quantities that are affected by this prior.

In principle, if the radius from the \citet{Chen(2017)} relation were inconsistent with the lower limit from the transit, using this relation could bias the mass of the 3.12-day planet. To ensure that was not the case, we used {\texttt{ExoFASTv2}} to run a 2-planet fit to only the RV data. We found that the resulting RV semi-amplitude of LTT~1445Ac is in agreement with both that from our {\texttt{ExoFASTv2}} global model fit with the radius constraint imposed, and the separate RV$+$GP fit we describe in \S \ref{subsec:cloutier}. 

Finally, we note that {\texttt{ExoFASTv2} rejects any samples where the planets' projected orbits cross into each other's Hill spheres, with the intention of constraining both planets' eccentricity by rejecting unstable configurations. While this is a reasonable approximation, it is not a substitute for a detailed analysis of the long-term stability of the system, which is beyond the scope of this work.}

\subsection{Confirmation of the 3.12-day Planet}\label{subsec:secondplanet} 

\begin{figure}

\hspace{-0.5cm}
\includegraphics[scale=.65,angle=0]{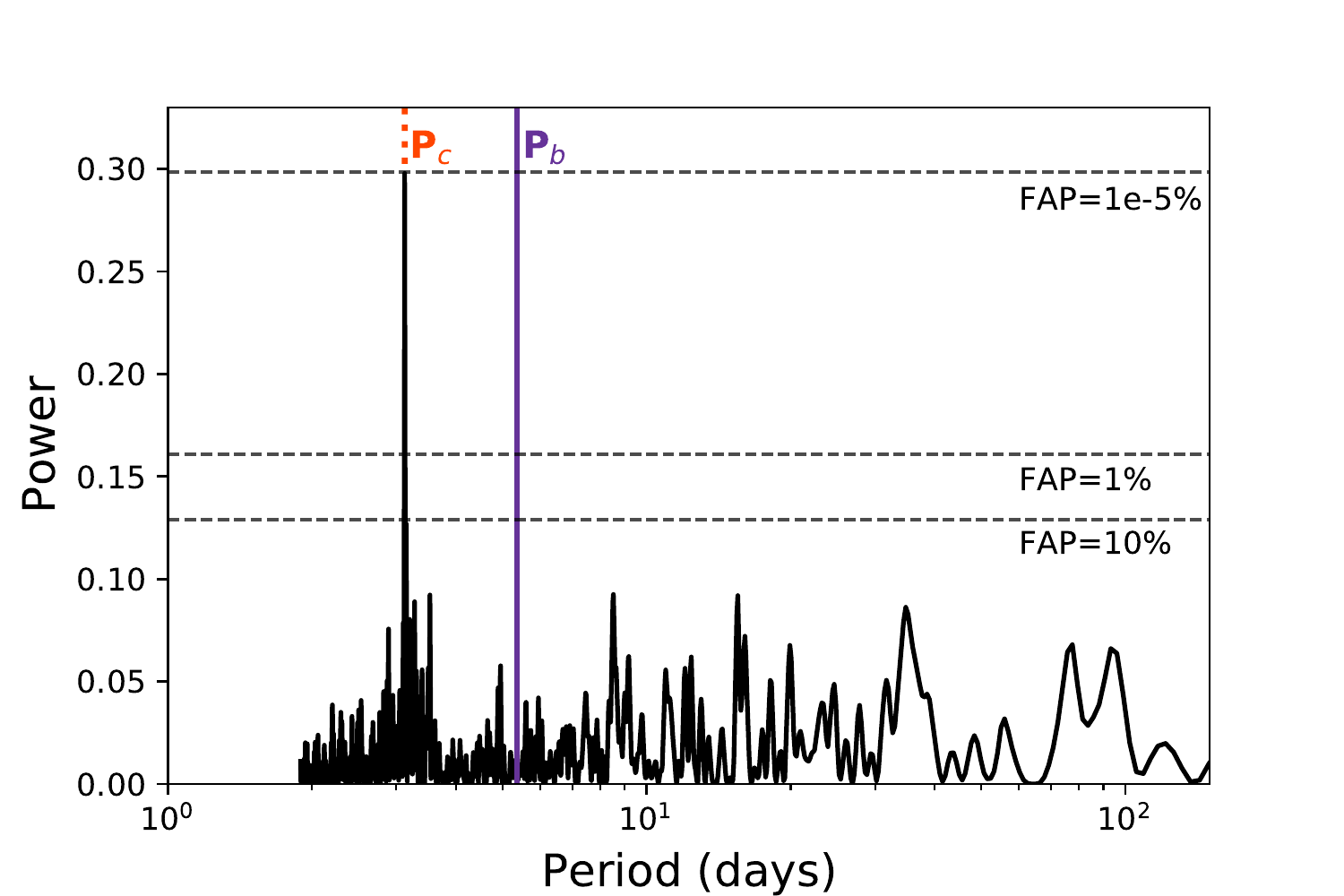}
\caption{Lomb-Scargle periodogram of the residuals from the 5.36-day planet fit to the RV data of LTT~1445A. False alarm probabilities of 10\%, 1\% and $1 \times 10^{-5}$\% are noted (black dashed lines), as are the orbital periods for the 3.12- and 5.36-day period planets (dotted and solid orange and purple lines, respectively). A highly significant peak at 3.12 days is evident, which confirms that the 3.12-day planet also orbits LTT~1445A. \label{fig:ls}}
\end{figure}

Because our only detection of the transits of the 3.12-day planet is from the \tess ~light curve data in which the light from all three stars is blended together, the light curve itself does not prove that the planet orbits star A. Instead the proof comes from our RVs of LTT~1445A: We first ran a global {\texttt{ExoFASTv2}} fit using both sectors of \tess ~data and all of the RVs, accounting for only the confirmed 5.36-day planet. We added the RV jitter output from the global model in quadrature with the uncertainties on the RV residuals to properly weight the uncertainties from each instrument. We then created a Lomb-Scargle (L-S) periodogram using the L-S tool \citep{VanderPlas(2015)} in {\texttt{astropy}} \citep{astropy(2013),astropy(2018)}, which we show in Figure \ref{fig:ls}. There is a peak at 3.12195 days with maximum L-S power at 0.298. The L-S False Alarm Probability (FAP) of a peak at this power is $1.06 \times 10^{-5}$\%, indicating high significance of the signal and indicating that indeed the 3.12-day planet also transits the primary star of LTT~1445.

We included the second planet in our global model to explore whether the addition of the second planet to our model would improve the fit to the data, under the assumption that it transits the primary star. The results indicate that the two-planet model is preferred over a single-planet model via the BIC and AIC metrics (BIC: $-21981$ vs. $-22209$; AIC: $-22169$ vs. $-22437$ for the one- and two-planet model, respectively). In \S \ref{subsec:cloutier} we conduct an independent analysis of the RV data using a Gaussian Process to account for stellar activity, and similarly find that the two-planet model is strongly preferred.

\begin{figure}
\includegraphics[scale=.50,angle=0]{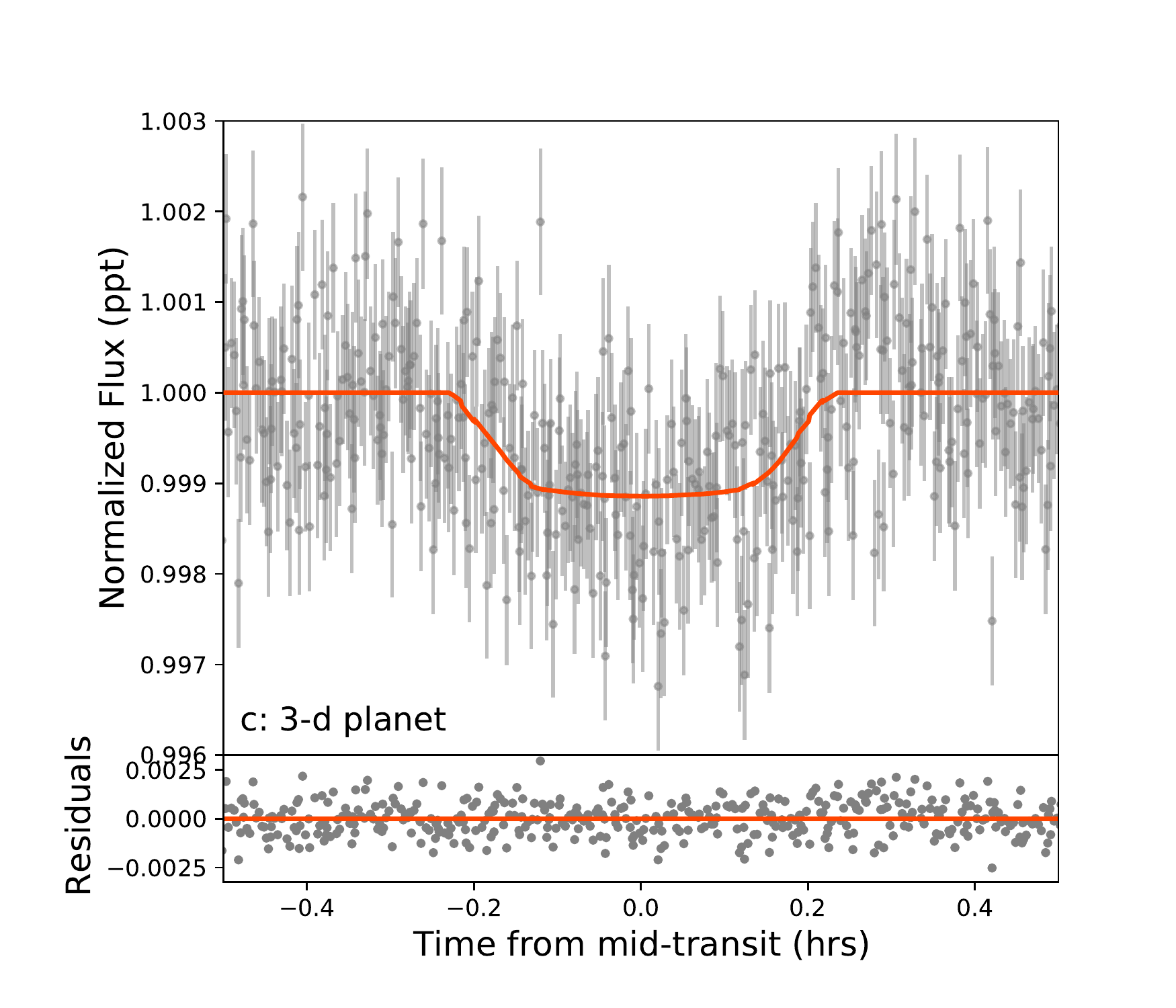}

\vspace{-0.4cm}

\includegraphics[scale=.50,angle=0]{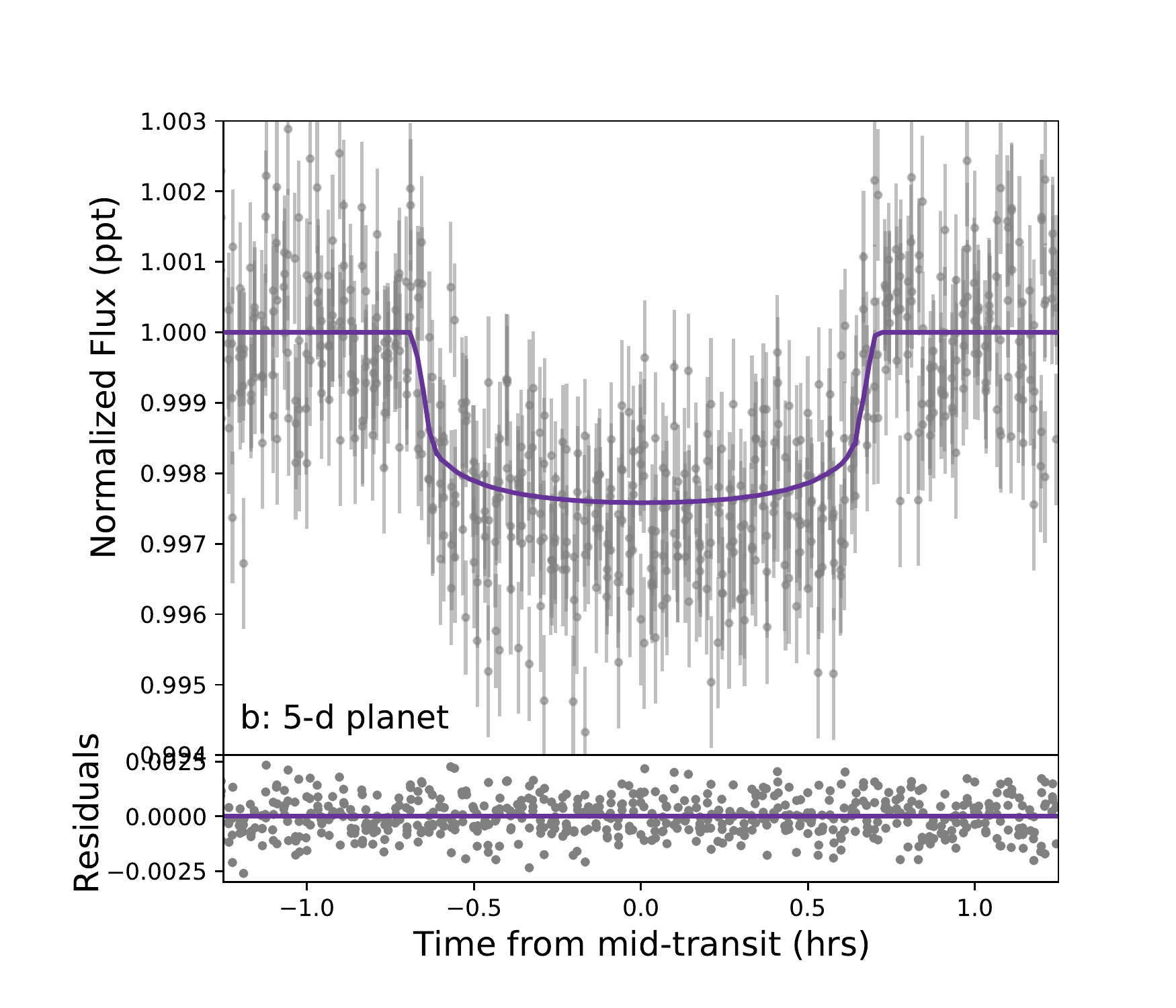}
\caption{Transits of LTT~1445Ac (top) and LTT~1445Ab (bottom) observed with \tess. The data have been deblended to remove any signal from the stellar BC components, and are shown in gray, while the orange and purple lines indicate the models for LTT~1445Ac and LTT~1445Ab. \label{fig:exofast_transits}}
\end{figure}

\begin{figure}

\includegraphics[scale=.50,angle=0]{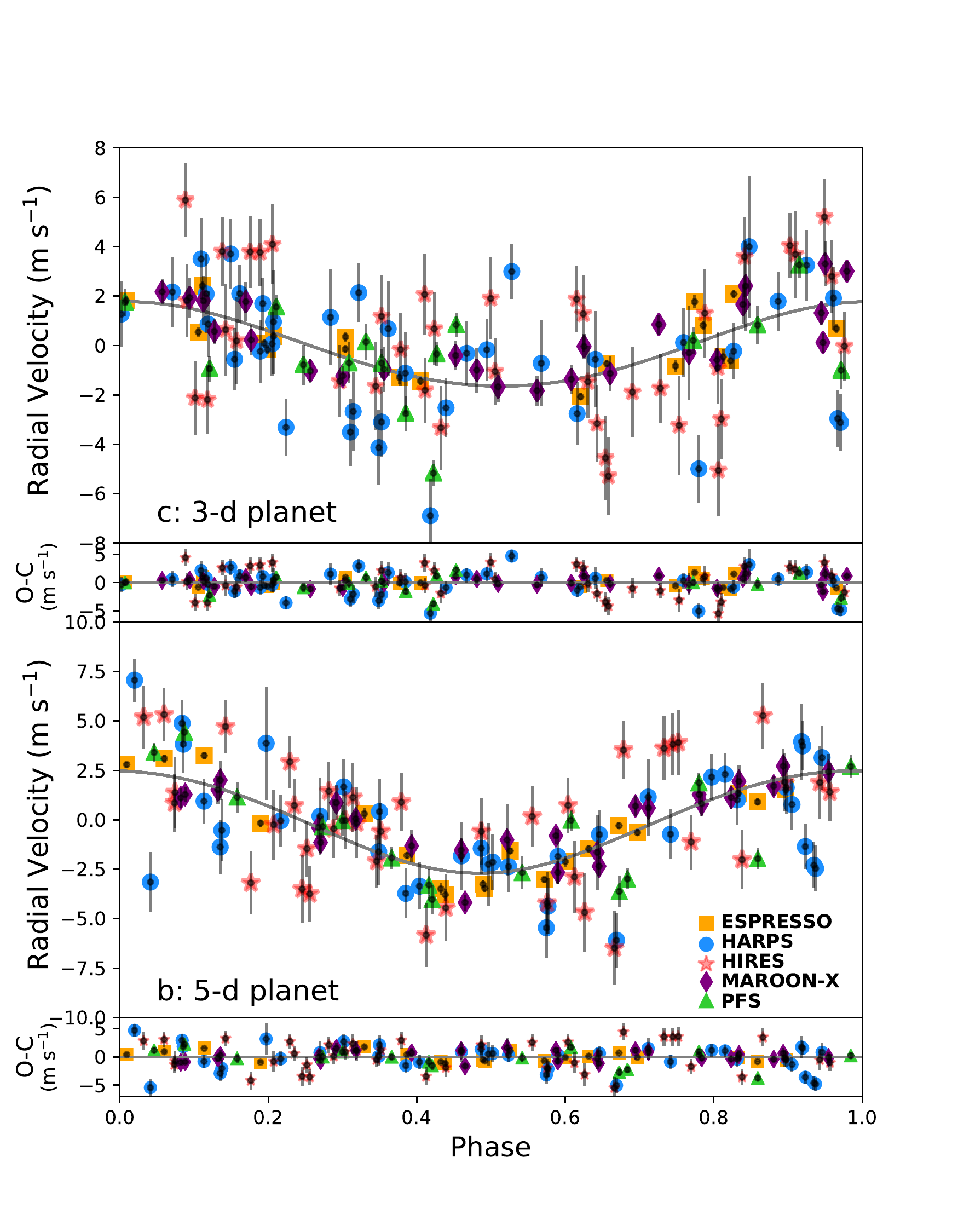}

\vspace{-0.55cm}

\includegraphics[scale=.55,angle=0]{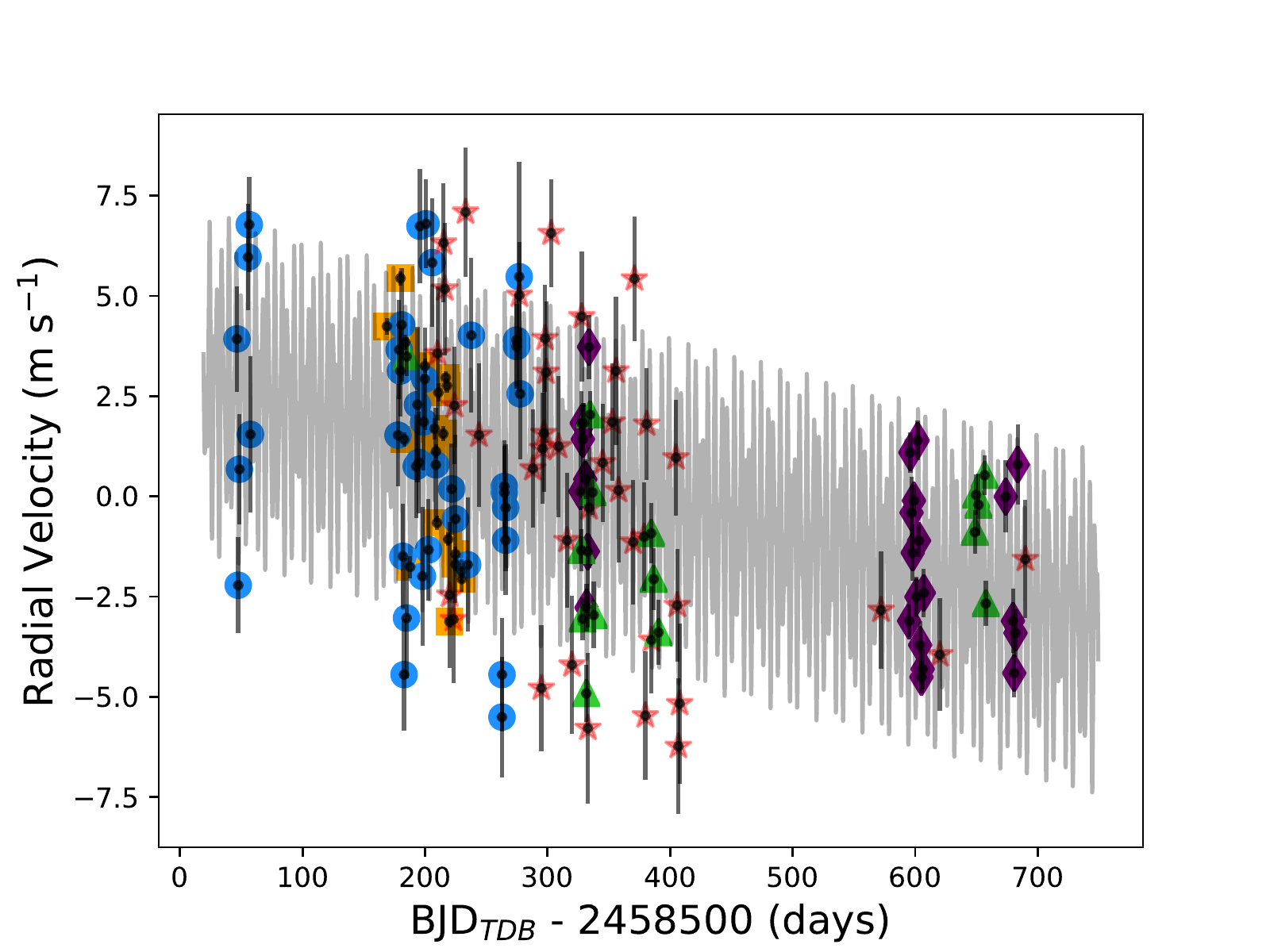}

\caption{RV orbital plots and residuals from the global {\texttt{ExoFASTv2}} 2-planet fit. In the top panel we show the RV data folded at the 3.12-day period (and corrected for the 5.36-day planet, and the slope), with the model for the 3.12-day planet overplotted in gray; residuals to the fit of both planets are shown underneath, phase-folded to the 3.12-day planet period. Similarly, in the third panel we show the RV data folded at the 5.36-day period (and corrected from the 3.12-day planet, and the slope), with the model for the 5.36-day overplotted in gray; residuals to the fit of both planets are shown underneath, phase-folded to the 5.36-day planet period. The data from ESPRESSO are plotted as orange squares, HARPS as blue circles, HIRES as pale red stars, MAROON-X as purple diamonds, and PFS as green triangles. The bottom panel illustrates the unphased RV data, where the gray curve indicates the combination of the 3.12-day and 5.36-day planet models and the linear slope due to the orbit of LTT~1445A about BC. \label{fig:rv_fits}}
\end{figure}

We thus adopt the two-planet model as the most appropriate description of the data and list the median values of the fit parameters from our global {\texttt{ExoFASTv2}} model for both planets in Table \ref{tab:master_table}. We show the resulting transit and RV plots for each planet in Figures \ref{fig:exofast_transits} and \ref{fig:rv_fits}.

\subsection{Radial Velocity Modeling with a Gaussian Process} \label{subsec:cloutier}

Active regions on the surface of a rotating star may produce non-orbital RV signals due to both the magnetic suppression of the net convective blueshift of the photosphere, and through the effect of spots downweighting portions of the stellar velocity field on the visible hemisphere of the star \citep{Meunier(2010)}. The temporal evolution of these activity signals can be detrimental to the recovery of accurate planetary signals if not properly accounted for. Here we assess the impact of stellar activity in our RV data by explicitly treating stellar activity with a GP regression model component \citep[e.g.,][]{Haywood(2014),Rajpaul(2015)}. Specifically, we model temporal correlations from rotationally-modulated magnetic activity using a quasi-periodic GP following the methodology of \cite{Cloutier(2017),Cloutier(2020a),Cloutier(2020b)}. To summarize, we adopt a separate quasi-periodic GP for each RV spectrograph due to their unique noise properties and the inherently chromatic nature of stellar activity. For each spectrograph indexed by $s$, the quasi-periodic covariance kernel is

\begin{equation}
  k_{sij} = a_s^2 \exp{\left[ -\frac{(t_i-t_j)^2}{2\lambda^2} - \Gamma^2 \sin^2{\left( \frac{\pi |t_i-t_j|}{P_{\rm GP}} \right)} \right]},
\end{equation}

\noindent where $a_s$ is the covariance amplitude for spectrograph $s$, $\lambda$ is the exponential timescale related to the lifetime of active regions, $\Gamma$ is the coherence parameter, and $P_{\rm GP}$ is the periodic timescale related to the stellar rotation. We also fit an additive scalar jitter $\sigma_{s,\text{RV}}$ for each spectrograph to account for any excess uncorrelated noise. The hyperparameters $\{ \lambda, \Gamma, P_{\rm GP} \}$ are all directly related to the active regions themselves and are therefore independent of the spectrograph. As such, these hyperparameters are shared among each spectrograph’s GP model.

We proceed with modeling our full RV dataset with a Keplerian model for each transiting planet, plus a GP activity model for each spectrograph. We use a custom-build \texttt{python} package, as described in \citet{Cloutier(2021)}, to sample the joint posterior of our complete RV model that incorporates the affine-invariant MCMC sampler \texttt{emcee} \citep{Foreman-Mackey(2013)} and the \texttt{george} package \citep{Ambikasaran(2014)} to evaluate the marginalized likelihood of our GP activity model components. We adopt the model parameter priors reported in Table~\ref{tab:lc_gp_priors}.
We recover the median a-posteriori GP hyperparameters and construct the predictive GP distributions for each spectrograph. We treat the mean function of each predictive distribution as the best-fit activity model for that spectrograph. 

\begin{deluxetable}{lccc}
\tabletypesize{\small}
\tablecaption{$RV+GP$ Priors  \label{tab:lc_gp_priors}}
\tablecolumns{2}
\tablenum{2}

\tablehead{
\colhead{Parameters} & 
\colhead{Prior} & 
}
\startdata
\multicolumn{2}{c}{\emph{GP hyper-parameters}} \\
$\rm{log}~a_{s}/\rm{(m/s)}$ & $\mathcal{U}(-5,5)$ \\
$\rm{log}~\lambda/\rm{days}$ & $\mathcal{U}(\log{1},\log{1000})$ \\
$\rm{log}~\Gamma$ & $\mathcal{U}(-3,3)$ \\
$P_{GP}$ [days] & $\mathcal{N}(85,22)$ \\
\multicolumn{2}{c}{\emph{LTT~1445Ab RV Parameters}} \\
$P_b$ [days]        & $\mathcal{N}(5.358766,0.000004)$ \\ 
$T_{0,b}$ [BJD]     & $\mathcal{N}(2458412.7085,0.0004)$ \\
$\rm{log}~K_b/\rm{(m/s)}$        & $\mathcal{U}(-3,3)$ \\
$h_b=\sqrt{e}~\cos{\omega}$        & $\mathcal{U}(-1,1)$ \\
$k_b=\sqrt{e}~\sin{\omega}$        & $\mathcal{U}(-1,1)$ \\
\multicolumn{2}{c}{\emph{LTT~1445Ac RV Parameters}} \\
$P_c$ [days]        & $\mathcal{N}(3.123904,0.000004)$ \\ 
$T_{0,c}$ [BJD]     & $\mathcal{N}(2458412.5816,0.0006)$ \\
$\rm{log}~K_c/\rm{(m/s)}$        & $\mathcal{U}(-3,3)$ \\
$h_c=\sqrt{e}~\cos{\omega}$        & $\mathcal{U}(-1,1)$ \\
$k_c=\sqrt{e}~\sin{\omega}$        & $\mathcal{U}(-1,1)$ \\
\multicolumn{2}{c}{\emph{Spectrograph Parameters}} \\
$\gamma_i$ [m/s] & $\mathcal{U}(-20,20) + \rm{median}(\textbf{rv})$ \\
$\rm{log}~s_i/(m/s)$ & $\mathcal{U}(-5,5)$ \\
\enddata
\end{deluxetable}



We use this analysis to evaluate the evidence for a second planet: Specifically, we calculate the Perrakis estimator \citep{Perrakis:2014}, as used by \citet{Diaz(2016)}, as an importance sampler to estimate the Bayesian model evidences for the one and two-planet RV+GP models. We find that the evidence ratio $\mathcal{Z}_2/\mathcal{Z}_1 = 1202$, which supports the confirmation of the 3.12-day planet in our RV time series \citep{Nelson:2020}. We conclude that a two-planet model is heavily favored over a one-planet model when including an explicit activity model in the form of a quasi-periodic GP. 


\subsection{Rotational Velocities and RVs of LTT~1445BC}
\label{subsec:bc_rot_per}

We have determined with our MEarth photometry that the 1.4-day rotational modulation we see in the \tess ~data originates from LTT~1445BC, as discussed in \S \ref{subsubsec:mearth_south_photometric_monitoring}. But we are not able to determine from the photometric time series alone whether the B or C component is the source of that signal. Very high resolution spectra of LTT~1445BC might permit the measurement of the rotational velocities of each component, which would allow us to determine the source. We noted in \S \ref{subsubsec:harps} that seven of our initial HARPS spectra of LTT~1445 were of the blended BC component. We downloaded these publicly available data and analyzed them using \texttt{todcor} \citep{Zucker(1994)} and the methods described in \citet{Winters(2020)} to extract individual RVs and rotational velocities ($v \sin i$) for the B and C components. 

We used an observation of Barnard's Star taken UT 2017 September 22 as the template for our analysis. We searched for the nonlinear least-squares maximum likelihood value for three parameters: flux ratio $\alpha$ and the two values of $v \sin i$. We calculate $\alpha$, the ratio of the flux of C relative to that of B, to be 0.31 at wavelengths $6760 - 6836$ \AA ~(echelle aperture 71 of the HARPS data), roughly equivalent to filter $R_{\rm KC}$. While this seems a bit low when compared to the $\alpha$ value of 0.50 we expect from photometry ($\Delta$ $R_{\rm KC}$ $= 0.75\pm0.03$ mag; \citealt{Henry(2006)}), the angular separation between B and C at the time of these observations is comparable to the HARPS 1\arcsec ~fiber diameter. Thus, it is reasonable that there is less light from LTT~1445C landing in the fiber. 


With this caveat in mind, we measure an average rotational broadening of 7.1 km \pers ~for LTT~1445C and 1.9 km \pers ~for LTT~1445B from the six spectra taken in September. These measurements suggest that the 1.4-day rotation period comes from the C component  and the 6.7-day marginal peak we see in the residual periodograms of the \tess ~and MEarth photometry in Figure \ref{fig:pgrams} is probably the rotation period of the B component. (We note that these results are not completely conclusive, due to $\sin i$ degeneracy: If the 6.7-day photometric period is spurious, then B could be the origin of the 1.4-day period, if we simply haven't detected the photometric period of C. But we view this possibility as unlikely.) Furthermore, we note that while we have not robustly detected rotational broadening in B, our solutions do slightly prefer some.

\begin{figure}
\includegraphics[scale=0.35,angle=0]{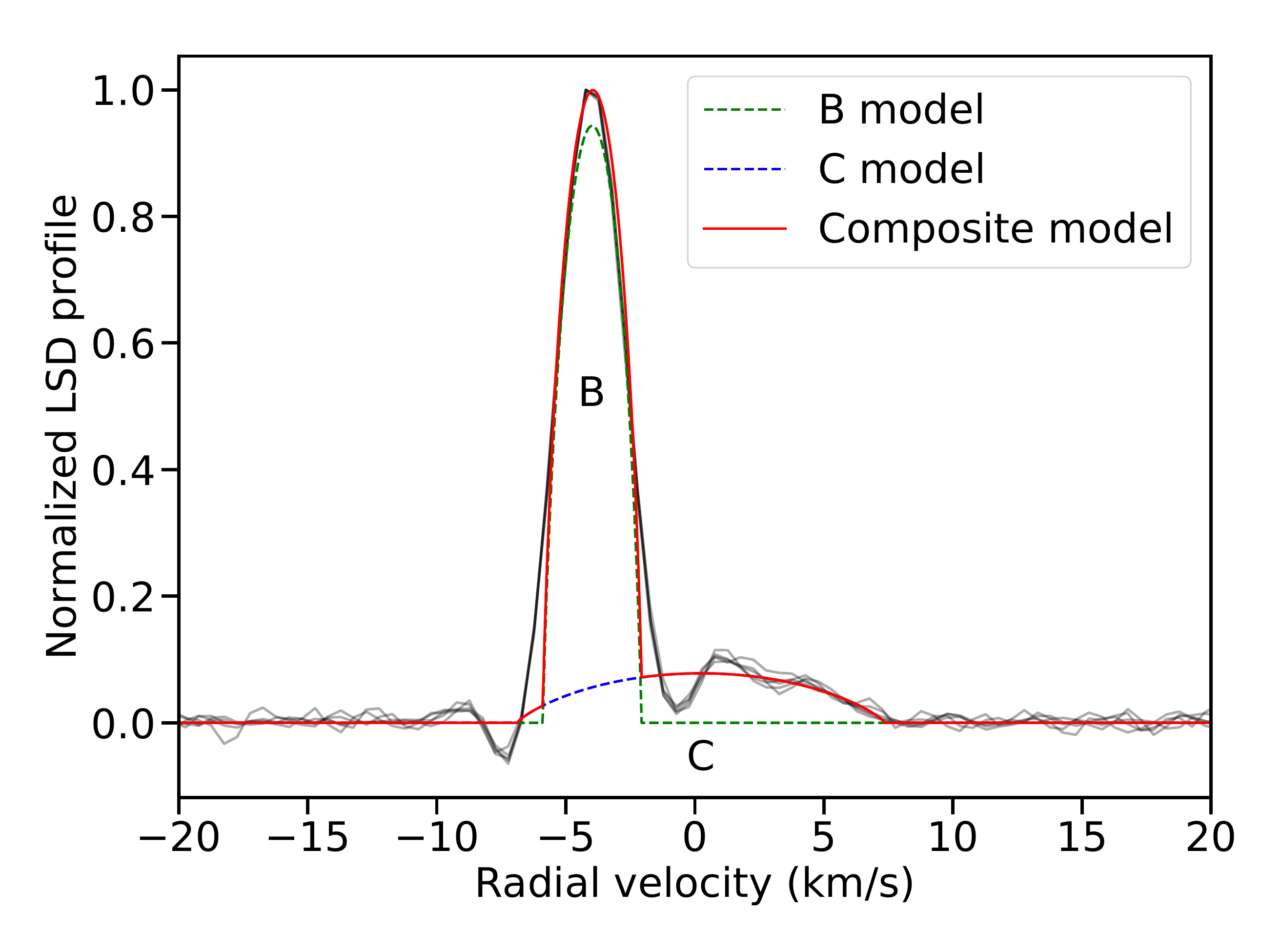}
\caption{Least-squares deconvolution profiles of LTT~1445B (green dashed line) and C (blue dashed line) from HARPS data (gray lines), which indicate the rotational broadening for each of the B and C components. The solid red line shows the joint fit of the broadened spectra.  \label{fig:bc_lsd}}
\end{figure}

As part of our analysis, we measure individual RVs for LTT~1445B and LTT~1445C of $-3.98\pm0.01$ km \pers ~and $0.24\pm0.20$ km \pers. We use the least-squares deconvolution \citep{Donati(1997)} to visualize the rotational broadening profiles of LTT~1445B and C in Figure \ref{fig:bc_lsd}.

\subsection{Check of Consistency Across Fits}

We performed a number of quality checks to ensure that our results are robust. We investigate three sets of parameters to check for consistency between model fits: We first compare the radius ratios we determine for each of the transit datasets. We then turn to the RV data, and compare both the RV semi-amplitude and the eccentricities from different analyses. 

We made separate adjustments to the PDCSAP light curves from both sectors of \tess ~data to correct the excess light from other stars in the optimal aperture. To ensure that we performed this correctly, we compare the results from fitting the transits of LTT~1445Ab from each sector of \tess ~data and find consistent planet-to-star radius ratios  $Rp/R_{*}$: From sector 4 data, we measure the radius ratio to be $0.0465\pm0.0015$, compared to that from sector 31, which is $0.0436^{+0.0015}_{-0.0014}$. We also compare the transit depth measurement results of LTT~1445Ab from our ground-based data to those from our global model. The measured radius ratio of the planet to the host star from ground-based data from MEarth and LCO are $0.0435\pm0.0021$ and $0.049^{+0.004}_{-0.003}$ (68.3\% interval), respectively. These are both in agreement with the measured $Rp/R_{*}$ of $0.0451^{+0.0014}_{-0.0013}$ from our global model. The agreement between all three measurements of the radius ratio of LTT~1445Ab indicates that we correctly adjusted the \tess ~dilution factor. We are not able to perform a radius ratio comparison for LTT~1445Ac because we do not yet have an independent ground-based detection of it.


We compare the results from our global model to the results from our modeling of the RVs with a GP. The measured RV semi-amplitudes from the RV$+$GP model for planets c and b are $1.49\pm0.17$ m \pers ~and $2.28\pm0.17$ m \pers, while they are $1.67\pm0.20$ m \pers, and $2.60\pm0.21$ m \pers ~from our global model for planets c and b, respectively. All measurements are in agreement. 

From the RV$+$GP model, the 95\% confidence intervals for the eccentricities of the orbits of c and b are $< 0.165$ and $< 0.089$, respectively. These are in agreement with the results from our global model, which are $<0.223$ and $<0.110$, respectively. 


Because the results from our RV$+$GP model and our global {\texttt{ExoFASTv2}} model are in agreement, we determined that the inclusion of a GP to model any periodic stellar variability in our RVs is unnecessary. Thus, we adopt the median values from our global {\texttt{ExoFASTv2}} model as our preferred results.

\begin{figure}
\centering
\includegraphics[scale=.50,angle=0]{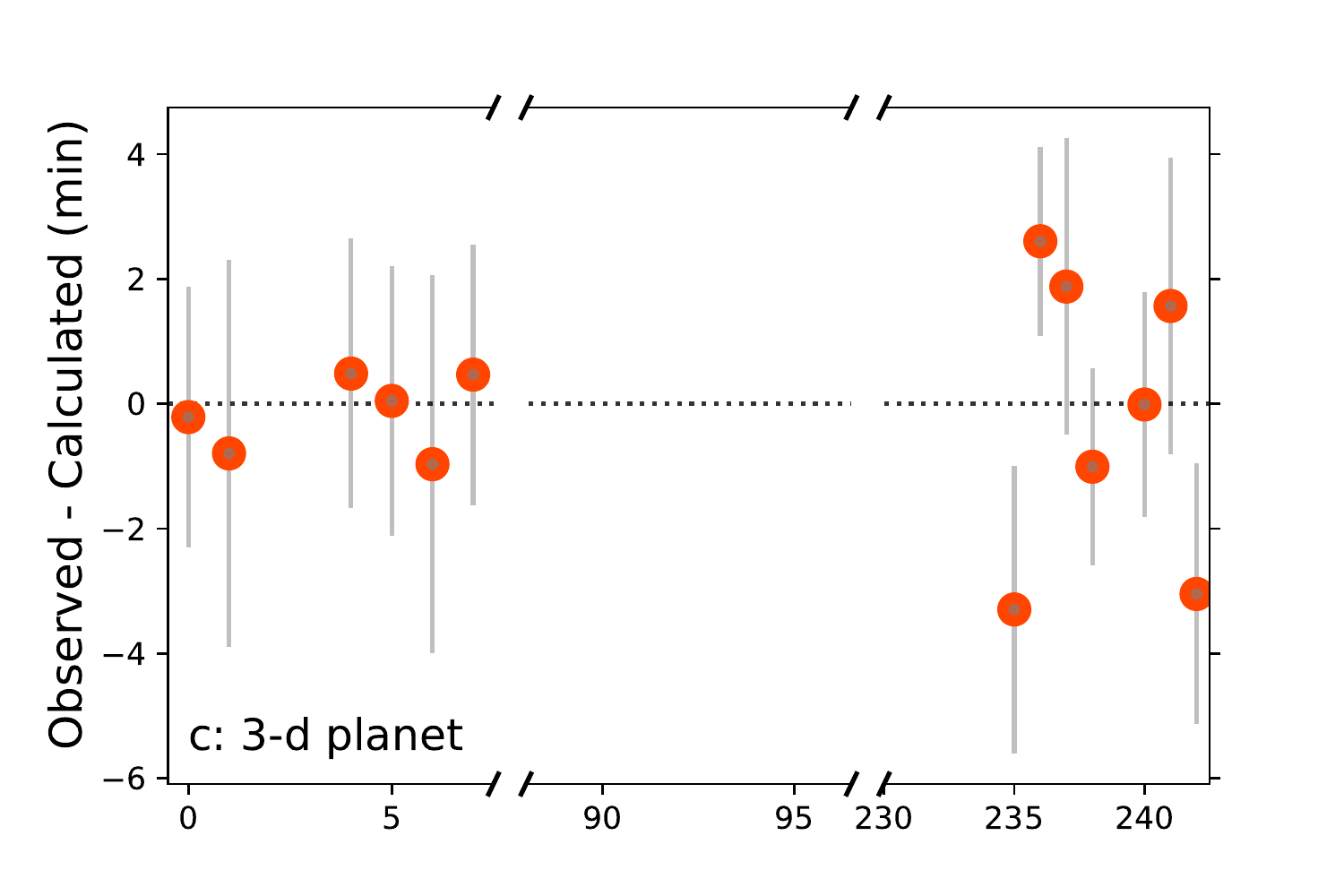}

\vspace{-0.4cm}

\includegraphics[scale=.50,angle=0]{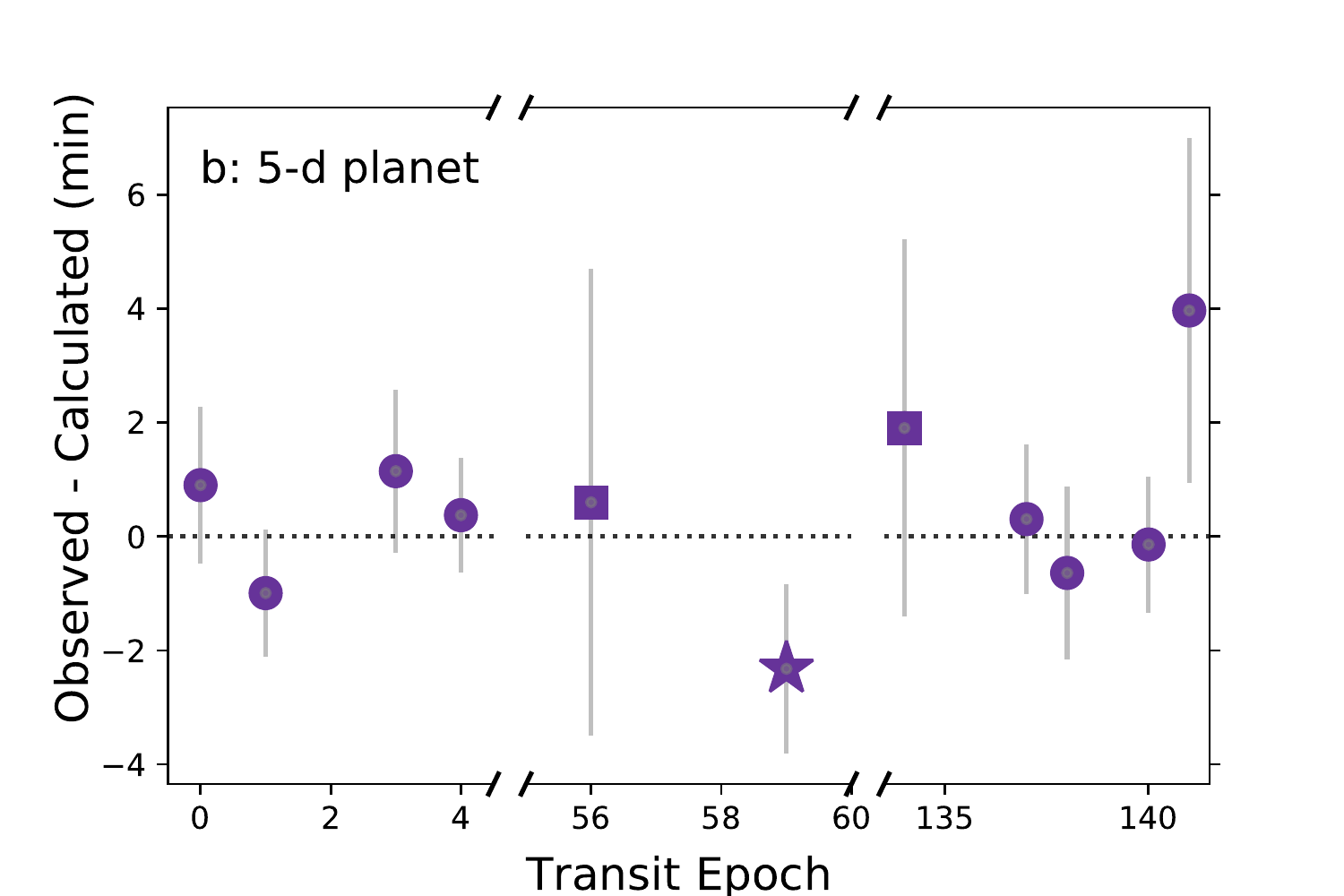}

\caption{Transit mid-times of the LTT~1445A planets. We show the measured transit mid-times for LTT~1445Ac (orange points, top panel) and for LTT~1445Ab (purple points, bottom panel). The transit mid-times for LTT~1445Ab from $TESS$, LCO, and MEarth data are noted as filled points, squares, and a star, respectively; transit mid-times for LTT~1445Ac are from TESS data only. As illustrated, the transit times for each planet are consistent with zero (dotted lines), when considering the uncertainties. \label{fig:ttvs}}
\end{figure}

\begin{figure}
    \centering
    \includegraphics[width=\hsize]{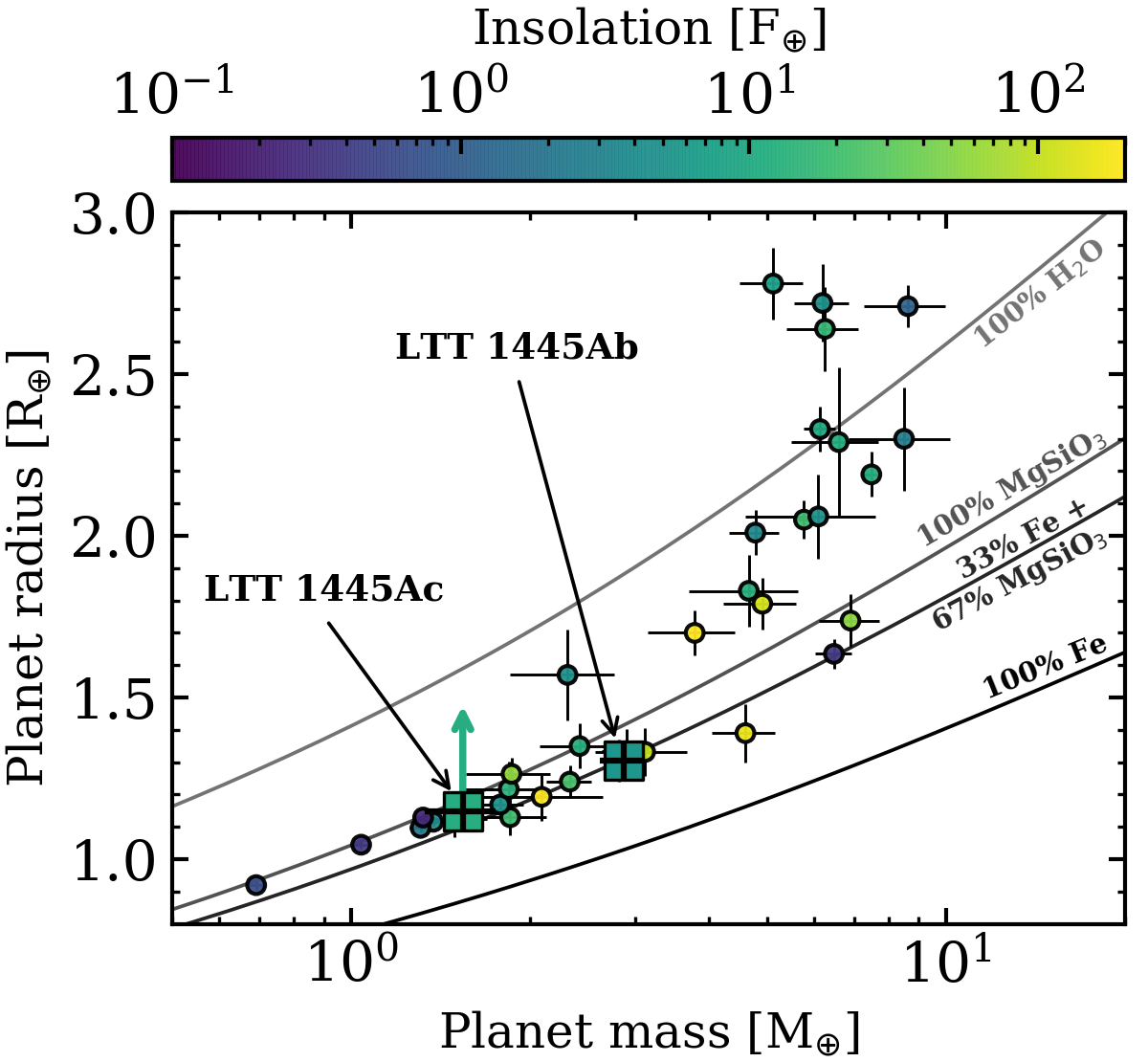}
    \caption{Masses and radii for small planets transiting M dwarfs and with precisely ($\geq 3\sigma$) measured masses. The solid curves are illustrative interior structure models of 100\% water, 100\% magnesium silicate rock, 33\% iron plus 67\% rock (i.e., Earth-like), and 100\% iron from  \citet{Zeng2013}. LTT~1445Ab and c are depicted by the square markers. LTT~1445Ab falls exactly on the composition curve corresponding to the Earth's ratio of iron (and nickel) to a magnesium-silicate. Although we measure the mass of LTT~1445Ac, we have deduced only a lower limit on its radius (green arrow), corresponding to the configuration where the transits are not grazing. If indeed they are not grazing, then the radius of LTT~1445Ac would be the minimum value depicted by the square marker here, and would lie on the same composition curve as LTT~1445Ab.}
    \label{fig:mr}
\end{figure}

\subsection{Search for Transit Timing Variations}

Multi-planet systems sometimes exhibit transit timing variations (TTVs) \citep{Agol(2005),Holman(2005)}, and we might expect them in the LTT~1445 system because the orbital periods of the two planets are not too dissimilar. Using the case of two planets not in resonance on circular orbits from \citet{Agol(2005)}, we calculate that TTVs for the LTT~1445A planets are expected to be on the order of 1 min for each planet. We utilized {\texttt ExoFASTv2}'s TTV capability to search for variations in each planet's transit mid-times, but did not detect any. We ran 213 chains for 3000 steps, thinned by 300, for a total number of 191.7 million simulations. For LTT~1445Ac, the shallow transit depth, grazing geometry, and low SNR of the light curves thwarted our efforts to reach a Gelman-Rubin value of less than 1.05 for the TTV parameters, except for LTT~1445Ac's transit epoch 239, which has two solutions. Therefore, for this TTV analysis only, we adopt the criterion of convergence at a Gelman-Rubin value less than 1.1 \citet{Gelman(2004)}. We omit transit 239 from our TTV analysis and provide the mid-transit times for each planet's transit from all of the time series photometry in Table \ref{tab:transit_times}. We show in Figure \ref{fig:ttvs} that the transit mid-times of both planets are consistent with a linear solution when considering the uncertainties.

\section{Discussion and Conclusions}
\label{sec:discuss}

We have measured the mass and radius of LTT~1445Ab to a precision of 9\% and 5\%, respectively. These precise mass and radius measurements permit the investigation of the planetary core mass fraction (cmf), which provides a probe for the planetary composition. Using the semi-empirical relation from \citet{Zeng(2016)}, we calculate a cmf of $0.425\pm0.283$ for LTT~1445Ab. This is illustrated in the radius vs. mass diagram in Figure \ref{fig:mr}, which includes composition curves from \citet{Zeng2013}. The planet falls squarely on the rocky compositional curve of 33\% iron plus 67\% magnesium-silicate. In the near future we hope to determine the abundances of key elements in the stellar photosphere, and see if they match the ratios we have inferred for the planet for its bulk composition.

We have detected a second planet transiting LTT~1445A. The presence of additional planets orbiting LTT~1445A is not surprising, given the previously known existence of LTT~1445Ab \citepalias{Winters(2019b)} and the recent result that $90^{+5}_{-21}$\% of mid-M dwarfs host multiple planets less massive than 10 \mearth ~and with orbital periods less than 50 days \citep{Cloutier(2021)}. LTT~1445A joins the growing list of mid-M dwarf systems in which additional planets have been revealed via follow-up observations of the original discovered planet. Some examples include GJ~357cd \citep{Luque(2019)}, GJ~1132c \citep{Bonfils(2018)}, GJ~3473c \citep{Kemmer(2020)}, and LHS~1140c \citep{Ment(2019)}.

While we have measured the mass of LTT~1445Ac to a precision of 13\%, the \tess ~data do not permit us to determine if all four points of contact are present in the transit curve of LTT~1445Ac, and hence whether the planetary transits are grazing. While the measured depth provides a minimum value for the radius of LTT~1445Ac, a grazing configuration allows acceptable solutions of arbitrarily large radii for the transiting object. If we impose a prior based on the mass-radius relation of \citet{Chen(2017)}, then we find an acceptable, non-grazing fit to the light curve, and the resulting planet radius is \pcradius ~\rearth. This would correspond to a cmf of $0.235\pm0.297$, again consistent with the telluric value. Ultimately this question requires a more precise light curve, and we eagerly await the observations scheduled with $HST$ (HST-GO-16503, PI: J.~Winters).

Intriguingly, our results indicate that the two planets have circular (or low-eccentricity) orbits that are misaligned by at least $-2.25_{-0.29}^{+0.28}$ degrees (as inferred from the posterior of the global model fit). This value is a lower limit for several reasons: First, it results from the inclusion of the mass-radius relation of \citet{Chen(2017)}, which results in non-grazing transits. Grazing transits by a larger planet would result in an even larger misalignment. Second, even if this solution is correct and the transits are not grazing, we observe only a projection onto the line-of-sight of the true misalignment between the planetary orbits. Third, there is an ($i, 180-i$) degeneracy and if the two orbits are on opposite sides of the central crossing, the mutual inclination would be $-2.91_{-0.45}^{+0.34}$ degrees, larger than the value presented above. For reference, the mutual inclination of the orbits of Venus and the Earth is 3.4 degrees.

We confirmed that the 1.4-day rotational modulation seen in the \tess ~data originates from LTT~1445BC. We have calculated the $v \sin i$ and RV values for LTT~1445B and C individually, and find that the 1.4-day signal is likely the rotation period of LTT~1445C, while the less statistically significant 6.7-day signal may be the rotation period of LTT~1445B. From the activity and mass of LTT~1445A, we expect its rotation period to be $85\pm22$ days, based on the empirical relation in \citet{Newton(2017)}. We have not yet detected the rotation period with MEarth, but our observations are ongoing. If our estimated rotation rates are correct, it implies that this triple star system is in a special evolutionary state, where A has spun down, B is in the process of spinning down, and C has not yet begun to spin down. LTT~1445 could be an important benchmark system for rotational evolution because it directly samples the mass dependence of the rapid spin-down phase discussed in the context of mid-to-late field M dwarfs by \cite{Newton(2017)}. Intriguingly, the dynamical masses of the stars B and C can eventually be estimated directly, although this may take some time given the moderately long orbital period of $36.2 \pm 5.3$ yrs \citepalias{Winters(2019b)}.

LTT~1445 is the second closest known transiting exoplanet system, and the closest one with an M-dwarf host star. 
Due to the efforts of MEarth \citep{Nutzman(2008),Irwin(2015)}, TRAPPIST \citep{Gillon(2016),Gillon(2017)}, and $TESS$ \citep{Ricker(2015)}, there are currently ten systems among the M dwarfs within 15 pc with masses $\leq 0.35~{\rm M}/\mdot$ that are known to host transiting planets: GJ~1214b \citep{Charbonneau(2009)}, GJ~1132bc \citep{Berta-Thompson(2015), Bonfils(2018)}, TRAPPIST-1 \citep{Gillon(2016),Gillon(2017),Grimm(2018)}, LHS~1140bc \citep{Dittmann(2017a), Ment(2019)}, LHS~3844b  \citep{Vanderspek(2019)}, LTT~1445Abc (\citetalias{Winters(2019b)}, this work), L~98-59bcd \citep{Kostov(2019), Cloutier(2019)}, GJ~357b \citep{Luque(2019)}, 2MA~0505-4756b \citep[TOI-540b;][]{Ment(2021)}, and GJ~486b \citep{trifonov21}. Although challenging, these planets are among the very best available for near-future studies of their atmospheres. \citet{Kempton(2018)} provide empirical relations that allow for a quantitative comparison of the SNR of spectra acquired with $JWST$, based on the stellar and planetary properties. For LTT~1445Ab, we calculate a transmission spectroscopy metric (TSM) of 30 for LTT~1445Ab and an emission spectroscopy metric (ESM) of 5.7. If LTT~1445Ac is indeed terrestrial with the values listed in Table \ref{tab:master_table}, then it would have an even higher TSM of 46 and an ESM of 8.7. For reference, \citet{Kempton(2018)} recommend minimum TSM and ESM values of 10 and 7.5, respectively, for planets that should be prioritized for future atmospheric studies. Furthermore, the proximity of LTT~1445A makes it a good candidate for stellar wind and mass-loss-rate measurements, as recently detailed in \citet{Wood(2021)}, to determine their effect on the planetary atmospheres, should those be present. 



\citetalias{Winters(2019b)} noted that, given the transit probability and number of nearby mid-to-late M dwarfs, \tess ~might find one more planet that is as accessible to follow-up study as LTT~1445Ab. LTT~1445Ac would appear to satisfy that criterion. As $TESS$ has observed roughly 80\% of the sky, the LTT~1445 system may remain the closest small star with transiting exoplanets. Then again, perhaps the remaining 20\% of the sky will hold some as-yet undiscovered jewels.

\startlongtable
\begin{deluxetable*}{lccccccc}
\tablecaption{ExoFASTv2 Median Values and 68\% Confidence Intervals for LTT~1445Ac \& LTT~1445Ab}
\tablecolumns{4}
\tablenum{3}
\tablehead{\colhead{Parameter} & \colhead{Units} &  \multicolumn{2}{c}{Values} }
\startdata
\multicolumn{2}{l}{Host Star Parameters:}&LTT~1445A\smallskip\\
~~~~$M_{*}$\dotfill & Stellar Mass (\mdot) & $0.257\pm0.014$\\
~~~~$R_{*}$\dotfill & Stellar Radius (R$_{\odot}$) &  $0.265^{+0.011}_{-0.010}$ \\
~~~~$T_{eff}$\dotfill & Stellar Effective Temp. (K) &  $3340\pm150$ \\
~~~~[Fe/H]\dotfill & Stellar Metallicity & $-0.34\pm0.09$\smallskip \\
~~~~$\dot{\gamma}$\dotfill &RV slope$^{1}$ (m \pers day$^{-1}$) & $-0.0081^{+0.0026}_{-0.0029}$ \\
\hline
\multicolumn{2}{l}{Planetary Parameters:}&LTT~1445Ac&LTT~1445Ab\smallskip\\
~~~~$P$\dotfill &Period (days)\dotfill & $3.1239035^{+0.0000034}_{-0.0000036}$&$5.3587657^{+0.0000043}_{-0.0000042}$\\
~~~~$R_P$\dotfill &Radius$^{6}$ (\re)\dotfill & $1.147^{+0.055}_{-0.054}$&$1.305^{+0.066}_{-0.061}$\\
~~~~$M_P$\dotfill &Mass (\me)\dotfill & $1.54^{+0.20}_{-0.19}$ & $2.87^{+0.26}_{-0.25}$ \\
~~~~$T_C$\dotfill &Time of conjunction$^{2}$ (\bjdtdb)\dotfill & $2458412.58159^{+0.00059}_{-0.00057}$&$2458412.70851^{+0.00040}_{-0.00039}$ \\
~~~~$T_T$\dotfill &Time of min. proj. sep.$^{3}$ (\bjdtdb)\dotfill & $2458412.58156^{+0.00059}_{-0.00057}$&$2458412.70851^{+0.00040}_{-0.00039}$\\
~~~~$i$\dotfill &Inclination$^{6}$ (Degrees)\dotfill & $87.43^{+0.18}_{-0.29}$&$89.68^{+0.22}_{-0.29}$ \\
~~~~$e$\dotfill &Eccentricity$^{4}$ \dotfill & $<0.223$&$<0.110$ \\
~~~~$K$\dotfill &RV semi-amplitude (m \pers)\dotfill & $1.67^{+0.21}_{-0.20}$ & $2.60\pm0.21$ \\
~~~~$\delta$\dotfill &Transit depth (fraction)\dotfill & $0.00157^{+0.00015}_{-0.00014}$&$0.00203\pm0.00012$ \\
~~~~$T_{14}$\dotfill &Total transit duration (days)\dotfill & $0.0201\pm0.0011$&$0.05697^{+0.00071}_{-0.00068}$ \\
~~~~$a$\dotfill &Semi-major axis (AU)\dotfill & $0.02661^{+0.00047}_{-0.00049}$&$0.03813^{+0.00068}_{-0.00070}$ \\
~~~~$R_P/R_*$\dotfill &Radius of planet in stellar radii$^{6}$ \dotfill & $0.0396^{+0.0018}_{-0.0017}$&$0.0451^{+0.0014}_{-0.0013}$ \\
~~~~$a/R_*$\dotfill &Semi-major axis in stellar radii \dotfill & $21.56^{+0.78}_{-0.82}$&$30.9^{+1.1}_{-1.2}$ \\
~~~~$b$\dotfill &Transit Impact parameter$^{6}$ \dotfill & $0.937^{+0.012}_{-0.016}$&$0.17^{+0.15}_{-0.12}$ \\
~~~~$b_S$\dotfill &Eclipse impact parameter$^{6}$ \dotfill &$0.977^{+0.20}_{-0.089}$&$0.17^{+0.14}_{-0.12}$\\
~~~~$\rho_P$\dotfill &Density$^{6}$ (g cm$^{-3}$)\dotfill & $5.57^{+0.68}_{-0.60}$&$7.1^{+1.2}_{-1.1}$\\
~~~~$log(g_P)$\dotfill &Surface gravity$^{6}$ \dotfill & $3.057^{+0.042}_{-0.043}$&$3.217^{+0.050}_{-0.053}$ \\
~~~~$\fave$\dotfill &Incident Flux (\fluxcgs)\dotfill & $0.0149^{+0.0032}_{-0.0027}$&$0.0073^{+0.0016}_{-0.0013}$ \\
~~~~$T_{eq}$\dotfill &Equilibrium temperature$^{5}$ (K)\dotfill & $508\pm25$&$424\pm21$\\
\smallskip\\\multicolumn{2}{l}{Wavelength Parameters:}&TESS\smallskip\\
~~~~$u_{1}$\dotfill &Linear limb-darkening coeff \dotfill & $0.156^{+0.079}_{-0.076}$\\
~~~~$u_{2}$\dotfill &Quadratic limb-darkening coeff \dotfill & $0.396\pm0.093$\\
~~~~$A_D$\dotfill &Dilution from neighboring stars \dotfill & $-0.012^{+0.048}_{-0.049}$\smallskip\\
\hline
\hline
\multicolumn{1}{l}{Spectrograph Parameters:}& & Relative RV Offset$^{1}$ & RV Jitter  \\
      & & (m \pers) & (m \pers)\smallskip\\
~~~~ESPRESSO \dotfill & &  $-5460.28^{+0.50}_{-0.53}$ & $0.96^{+0.23}_{-0.17}$   \\
~~~~HARPS (2019) \dotfill & & $-5458.7^{+1.4}_{-1.5}$ & $1.9^{+1.7}_{-1.0}$  \\
~~~~HARPS (2020) \dotfill & &  $-5453.51^{+0.63}_{-0.64}$ & $2.30^{+0.45}_{-0.38}$  \\
~~~~HIRES \dotfill & & $-1.75\pm0.45$ & $2.29^{+0.45}_{-0.39}$ \\
~~~~MAROON-X (2019) \dotfill & & $-0.14^{+0.58}_{-0.53}$  & $1.16^{+0.74}_{-0.49}$ \\
~~~~MAROON-X (2020) \dotfill & &  $1.84^{+0.78}_{-0.72}$ & $0.93^{+0.29}_{-0.22}$  \\
~~~~PFS \dotfill & & $0.41^{+0.58}_{-0.54}$ & $1.87^{+0.60}_{-0.43}$\smallskip\\
\hline
\hline
\multicolumn{1}{l}{$TESS$ Transit Parameters:}& &  Added Variance $\sigma^{2}$ & Baseline flux $F_0$\smallskip\\ 
~~~~2018\dotfill & &  $0.000000170^{+0.000000039}_{-0.000000036}$ & $0.999954\pm0.000031$ \\
~~~~2020\dotfill& &  $-0.000000009^{+0.000000038}_{-0.000000036}$ & $1.000084\pm0.000030$ \\
\enddata
\label{tab:master_table}
\tablenotetext{}{See Table 3 in \citet{Eastman(2019)} for a detailed description of all parameters.}
\tablenotetext{1}{Reference epoch = 2458868.217394}
\tablenotetext{2}{Time of conjunction is commonly reported as the "transit time."}
\tablenotetext{3}{Time of minimum projected separation is a more correct "transit time."}
\tablenotetext{4}{$2-\sigma$ (95\%) upper limits.}
\tablenotetext{5}{Assumes no albedo and perfect redistribution.}
\tablenotetext{6}{The value for planet c was derived using the measured mass, the lower limit on the radius from the light curve, and \citet{Chen(2017)} exoplanet mass-radius relation to estimate the planetary radius.}
\end{deluxetable*}

\startlongtable
\begin{deluxetable*}{lccccc}
\tablecaption{Median values and 68\% confidence interval for planet transit mid-times}
\tablenum{4}
\tablehead{\colhead{Transit} & \colhead{Planet} & \colhead{Epoch} & \colhead{$T_T$} }
\startdata
TESS UT 2018-10-21 (TESS) & c & 0 & $2458412.5815^{+0.0014}_{-0.0015}$ \\
TESS UT 2018-10-24 (TESS) & c & 1 & $2458415.7050^{+0.0023}_{-0.0020}$ \\
TESS UT 2018-11-02 (TESS) & c & 4 & $2458425.0776 \pm 0.0015$ \\
TESS UT 2018-11-05 (TESS) & c & 5 & $2458428.2012^{+0.0014}_{-0.0016}$ \\
TESS UT 2018-11-08 (TESS) & c & 6 & $2458431.3244^{+0.0024}_{-0.0018}$ \\
TESS UT 2018-11-11 (TESS) & c & 7 & $2458434.4493^{+0.0015}_{-0.0014}$ \\
TESS UT 2020-10-24 (TESS) & c & 235 & $2459146.6968^{+0.0017}_{-0.0015}$ \\
TESS UT 2020-10-27 (TESS) & c & 236 & $2459149.8248^{+0.0010}_{-0.0011}$ \\
TESS UT 2020-10-30 (TESS) & c & 237 & $2459152.9482^{+0.0018}_{-0.0015}$ \\
TESS UT 2020-11-02 (TESS) & c & 238 & $2459156.0701 \pm 0.0011$ \\
TESS UT 2020-11-08 (TESS) & c & 240 & $2459162.3186^{+0.0013}_{-0.0012}$ \\
TESS UT 2020-11-11 (TESS) & c & 241 & $2459165.4436^{+0.0017}_{-0.0016}$ \\
TESS UT 2020-11-15 (TESS) & c & 242 & $2459168.5643^{+0.0015}_{-0.0014}$ \\
\hline
TESS UT 2018-10-21 (TESS) & b & 0 & $2458412.70898^{+0.00099}_{-0.00093}$ \\
TESS UT 2018-10-26 (TESS) & b & 1 & $2458418.06643^{+0.00080}_{-0.00076}$ \\
TESS UT 2018-11-06 (TESS) & b & 3 & $2458428.78545^{+0.00099}_{-0.0010}$ \\
TESS UT 2018-11-11 (TESS) & b & 4 & $2458434.14368^{+0.00075}_{-0.00065}$ \\
LCO UT 2019-08-16 (z') & b & 56 & $2458712.7997^{+0.0030}_{-0.0027}$ \\
MEarth UT 2019-09-02 (TESS) & b & 59 & $2458728.87397^{+0.00096}_{-0.0011}$ \\
LCO UT 2020-10-08 (z') & b & 134 & $2459130.7844^{+0.0018}_{-0.0028}$ \\
TESS UT 2020-10-24 (TESS) & b & 137 & $2459146.85959^{+0.00092}_{-0.00090}$ \\
TESS UT 2020-10-29 (TESS) & b & 138 & $2459152.21770^{+0.00080}_{-0.0013}$ \\
TESS UT 2020-11-09 (TESS) & b & 140 & $2459162.93558^{+0.00089}_{-0.00077}$ \\
TESS UT 2020-11-14 (TESS) & b & 141 & $2459168.2972^{+0.0019}_{-0.0023}$ \\
\enddata
\label{tab:transit_times}
\end{deluxetable*}

\facilities{{TESS}, {MEarth}, {LCOGT}, {ESO:3.6m (HARPS)}, {Gemini:Gillett (MAROON-X)}, {Keck (HIRES)}, {Magellan:Clay (PFS)}, {VLT (ESPRESSO)}}

\software{{\texttt{AstroImageJ}} \citep{Collins:2017}, {\texttt{astropy}} \citep{astropy(2013),astropy(2018)}, {\texttt {celerite}} \citep{Foreman-Mackey(2017)}, {\texttt{ExoFASTv2}} \citep{Eastman(2013),Eastman(2017)}, {\texttt {exoplanet}} \citep{exoplanet:exoplanet}, \texttt{george} \citep{Ambikasaran(2014)}, IDL, IRAF, {\texttt{LcTools II}} \citep{Schmitt(2021)}, {\texttt{PYMC3}} \citep{exoplanet:pymc3}, {\texttt{python}}, {\texttt{serval}} \citep{Zechmeister(2018)}, {\texttt{starry}} \citep{exoplanet:luger18}, {\texttt{TAPIR}} \citep{Jensen:2013}, Time Utilities \citep{Eastman(2010)}, \texttt{todcor} \citep{Zucker(1994)}}.

\vspace{5mm}
\begin{center}
\large
Acknowledgments
\end{center}
\normalsize

We thank the referee for a thoughtful and prompt review that improved the manuscript. This work is made possible by a grant from the John Templeton Foundation. The opinions expressed in this publication are those of the authors and do not necessarily reflect the views of the John Templeton Foundation. The MEarth Team gratefully acknowledges funding from the David and Lucile Packard Fellowship for Science and Engineering (awarded to D.C.). This material is based upon work supported by the National Science Foundation under grant AST-1616624, and work supported by the National Aeronautics and Space Administration under Grant No.~80NSSC18K0476 issued through the XRP Program.

This paper includes data collected by the TESS mission that are publicly available from the Mikulski Archive for Space Telescopes (MAST). We acknowledge the use of public TESS data from pipelines at the TESS Science Office and at the TESS Science Processing Operations Center. This research has made use of the Exoplanet Follow-up Observation Program website, which is operated by the California Institute of Technology, under contract with the National Aeronautics and Space Administration under the Exoplanet Exploration Program. Resources supporting this work were provided by the NASA High-End Computing (HEC) Program through the NASA Advanced Supercomputing (NAS) Division at Ames Research Center for the production of the SPOC data products. Funding for the $TESS$ mission is provided by the NASA's Science Mission Directorate. 

This work makes use of observations from the LCOGT network.
Part of the LCOGT telescope time was granted by NOIRLab through the Mid-Scale Innovations Program (MSIP). MSIP is funded by NSF. This work was enabled by observations made from the Gemini-North telescope, located within the Maunakea Science Reserve and adjacent to the summit of Maunakea. The authors wish to recognize and acknowledge the very significant cultural role and reverence that the summit of Maunakea has always had within the indigenous Hawaiian community. We are grateful for the privilege of observing the Universe from a place that is unique in both its astronomical quality and its cultural significance. The international Gemini Observatory, a program of NSF's NOIRLab, is managed by the Association of Universities for Research in Astronomy (AURA) under a cooperative agreement with the National Science Foundation. on behalf of the Gemini Observatory partnership: the National Science Foundation (United States), National Research Council (Canada), Agencia Nacional de Investigaci\'{o}n y Desarrollo (Chile), Ministerio de Ciencia, Tecnolog\'{i}a e Innovaci\'{o}n (Argentina), Minist\'{e}rio da Ci\^{e}ncia, Tecnologia, Inova\c{c}\~{o}es e Comunica\c{c}\~{o}es (Brazil), and Korea Astronomy and Space Science Institute (Republic of Korea). The MAROON-X spectrograph was funded by the David and Lucile Packard Foundation, the Heising-Simons Foundation, the Gemini Observatory, and the University of Chicago. This paper includes data gathered with the 6.5 meter Magellan Telescopes located at Las Campanas Observatory, Chile. Some of the data presented herein were obtained at the W. M. Keck Observatory, which is operated as a scientific partnership among the California Institute of Technology, the University of California and the National Aeronautics and Space Administration. The Observatory was made possible by the generous financial support of the W. M. Keck Foundation.

This work has made use of data from the European Space Agency (ESA) mission
{\it Gaia} (\url{https://www.cosmos.esa.int/gaia}), processed by the {\it Gaia}
Data Processing and Analysis Consortium (DPAC,
\url{https://www.cosmos.esa.int/web/gaia/dpac/consortium}). Funding for the DPAC
has been provided by national institutions, in particular the institutions
participating in the {\it Gaia} Multilateral Agreement. Data products from the Two Micron All Sky Survey, which is a joint
project of the University of Massachusetts and the Infrared Processing
and Analysis Center/California Institute of Technology, funded by the
National Aeronautics and Space Administration (NASA) and the NSF have
been used in this publication. This work has made use of the Washington Double Star Catalog maintained at the U.S. Naval Observatory. This work has
made use of the Smithsonian Astrophysical Observatory/NASA
Astrophysics Data System. 

R.C. acknowledges support from the Banting Postdoctoral Fellowship program, administered by the Government of Canada. N.A.-D. acknowledges the support of FONDECYT project 3180063. J.M.A.M. is supported by the NSF Graduate Research Fellowship, grant No. DGE-1842400. J.M.A.M. also acknowledges the LSSTC Data Science Fellowship Program, which is funded by LSSTC, NSF Cybertraining Grant No. 1829740, the Brinson Foundation, and the Moore Foundation; his participation in the program has benefited this work.  D.H. acknowledges support from the Alfred P. Sloan Foundation, the National Aeronautics and Space Administration (80NSSC21K0652), and the National Science Foundation (AST-1717000). K.H. acknowledges support from STFC grant ST/R000824/1. T.F. acknowledges support from the University of California President's Postdoctoral Fellowship Program. P.D. acknowledges support from a National Science Foundation Astronomy and Astrophysics Postdoctoral Fellowship under award AST-1903811. 

\bibliographystyle{aasjournal}
\bibliography{masterref.bib}

\end{document}